\documentclass{article}

\usepackage{arxiv}

\usepackage[utf8]{inputenc} 
\usepackage[T1]{fontenc}    
\usepackage{hyperref}       
\usepackage{url}            
\usepackage{booktabs}       
\usepackage{amsfonts}       
\usepackage{nicefrac}       
\usepackage{microtype}      
\usepackage{lipsum}
\usepackage{fancyhdr}       
\usepackage{graphicx}       
\graphicspath{{media/}}     
\usepackage{xcolor}
\usepackage{subcaption}
\usepackage{float}
\usepackage{wrapfig}
\usepackage{amsmath}

\title{Active Learning on Neural Networks through Interactive Generation of Digit Patterns and Visual Representation
}

\author{
  Dong H. Jeong \\
  Department of Computer Science and Information Technology \\
  University of the District of Columbia \\
  Washington\\
  \texttt{djeong@udc.edu} \\
   \And
  Jin-Hee Cho \\
  Department of Computer Science \\
  Virginia Tech. \\
  Falls Church\\
  \texttt{jicho@vt.edu} \\
  \AND
  Feng Chen \\
  Department of Computer Science \\
  University of Texas at Dallas \\
  Richardson \\
  \texttt{feng.chen@utdallas.edu} \\
  \And
  Audun J{\o}sang \\
  Department of Informatics \\
  University of Oslo \\
  Oslo \\
  \texttt{audun.josang@mn.uio.no} \\
  \And
  Soo-Yeon Ji \\
  Department of Computer Science \\
  Bowie State University \\
  Bowie\\
  \texttt{sji@bowiestate.edu} \\
}

\begin{document}
\maketitle

\begin{abstract}
Artificial neural networks (ANNs) have been broadly utilized to analyze various data and solve different domain problems. However, neural networks (NNs) have been considered a black box operation for years because their underlying computation and meaning are hidden. Due to this nature, users often face difficulties in interpreting the underlying mechanism of the NNs and the benefits of using them. In this paper, to improve users' learning and understanding of NNs, an interactive learning system is designed to create digit patterns and recognize them in real time. To help users clearly understand the visual differences of digit patterns (i.e., 0 $\sim$ 9) and their results with an NN, integrating visualization is considered to present all digit patterns in a two-dimensional display space with supporting multiple user interactions. An evaluation with multiple datasets is conducted to determine its usability for active learning. In addition, informal user testing is managed during a summer workshop by asking the workshop participants to use the system. 
\end{abstract}


\section{Introduction}
\label{Introduction}

Artificial Intelligence (AI) has significant impacts in many disciplines, such as medicine \cite{miller2018artificial}, education \cite{du2016recent}, and earthquake \cite{karasozen2020earthquake}, and so forth. AI enables machines to learn data to perform various tasks to function like humans \cite{duan2019artificial}. Examples of AI-enabled technologies include abnormal behavior detection in banking transactions, home protection using camera-based monitoring and object recognition, customized learning mechanisms providing tailored lessons to students, and driving assistance with upgraded and autonomous driving technologies. AI also has received much attention as a tool in education to improve students' learning and knowledge acquisition  \cite{Popenici2017}. 

Despite the powerful capability of AI and its numerous applications in our daily lives, AI has been known as a black box tool because it is not easy to understand the logic behind its internal computations \cite{wang2020should}. This motivated researchers to design trustworthy and interpretable AI systems \cite{BARREDOARRIETA202082,HOLZINGER2022263}. However, it is still difficult to understand AI systems due to the well-known issue of transparency \cite{Haresamudram2023} because machine or deep learning algorithms used in AI models are highly complex. For instance, neural networks (NNs) or deep NNs (DNNs) often include thousands of artificial neurons to learn from and process large amounts of data. Because of the numerous neurons and their complex interconnections, it is difficult to determine how decisions are made \cite{Bathaee2018TheAI}. Understanding how the AI models work and generate resulting predictions is a critical step in interpreting the meaning of AI's outcomes. But, it remains challenging to understand data using a predictive model that finds patterns from training the data and analyzes the difference between predictions and the patterns. 

This paper aims to integrate AI technologies into learning through hands-on practices. Specifically, designing an interactive learning system is considered to assist users in understanding NNs clearly through multiple learning activities. To support an interactive learning environment on NNs, we have designed the system with a pattern generator, where a user can generate digit patterns. In addition, providing a visual representation of the patterns is considered to help the user understand the differences among the generated patterns.  Whenever the user creates patterns, the user can experience NN training and recognition in real time. To represent the patterns in a 2D display space, Principal Component Analysis (PCA) was used to reduce the dimensions of the patterns and present them in a lower-dimensional space \cite{roweis1997algorithms, Jeong09}. By interactively navigating the display space, the user can identify the similarities and differences between the patterns. To evaluate the system's effectiveness for active learning, 2400-digit patterns are generated and used to test the system. A broadly known handwritten digits dataset (called MNIST) is also used to determine the capability of supporting a real-time interactive digits analysis. To understand the usefulness of the system, informal user testing was organized during a summer workshop by asking the workshop participants to use the system. We found that they understood NNs well by initiating active learning with the system.

The rest of this paper is structured in six sections. Section~\ref{sec:Related Work} provides previous studies on designing educational systems for understanding AI. In~Section~\ref{sec:System Design}, the designed system is explained. Section~\ref{sec:System Design} includes a detailed explanation about the applied neural networks and visual representation. Section~\ref{sec:Evaluating the Effectiveness of the System} shows the conducted evaluation of interactive learning on recognizing digits in real time. After discussing interesting insights in Section~\ref{sec:Discussion}, we conclude this paper by providing possible future work in Section~\ref{sec:Conclusion}.


\section{Related Work}
\label{sec:Related Work}

Due to high interest in AI,  most education institutes, including colleges or high schools, have introduced new AI degree programs \cite{tyson2007science}.  AI has become a powerful paradigm in scientific research communities due to its diverse applications in broad and various domains \cite{XU2021100179}. Due to this popularity, many students have shown a strong interest in understanding AI. In particular, they have exposed their high interest in deep learning (DL) because it has been commonly used to detect complex patterns in high-dimensional data with little or no human interventions. However, understanding the underlying ideas of the output prediction in DL is not trivial due to the black-box nature of the AI models \cite{you2019real,lamy2019visual, Guidotti2018}. Li and et al. \cite{Hao18} explored various visualization techniques to understand the structure of neural loss functions and their effectiveness. Chatzimparmpas and et al. \cite{chatzimparmpas2020survey}  emphasized how important information visualization is in understanding machine learning (ML) models and enhancing trust in ML. Although they highlighted the importance of utilizing visualization in ML, their primary considerations fell into addressing specific domain problems instead of helping students understand the internal computation of ML.

Computer science education researchers have developed various tools to improve students' knowledge of AI technologies. Mariescu-Istodor and Jormanainen \cite{mariescu2019machine} developed a web-based tool for high school students to enhance their knowledge in recognizing objects using ML. They designed the tool to identify objects using a camera and determine their object classes in real time based on training samples. In this tool, when a student gives a wrong answer, the student sees a question mark rather than a message saying the answer is wrong. If an object has been misclassified (i.e., the student says a wrong answer), the tool could fix such a mistake by correctly training and classifying its class name with additional samples. The authors aimed to design the tool to motivate students by improving their class engagement. You and Yin \cite{you2019real} developed a device (called Omega) to enhance college students' understanding on NNs by addressing the black box nature and representing their interactions during NN training steps. In particular, the device visually presented the weight changes in hidden layers during the NN training. Lamy and Tsopra \cite{lamy2019visual} introduced a visual translation of simple NNs to prove the visual interpretation using rainbow boxes with adding interactive functionality. Kim and Shim \cite{kim2022development} emphasized the need of providing AI education for non-engineering major students by creating a  visual solution. Although numerous studies have been conducted to design practical approaches to improve students' learning, most studies mainly aimed to teach users NN training steps.

Unlike the existing approaches explained above, our study differs in that the proposed interactive learning system enables users to create input data patterns and train NNs interactively. This will significantly increase the user's learning and knowledge gained on NNs because it supports real-time computation and recognition of the user-generated digit patterns.  

\section{System Design}
\label{sec:System Design}

\begin{figure}[H]
    \centering
    \begin{subfigure}[b]{.64\linewidth}
        \includegraphics[height=2in, width=\linewidth]{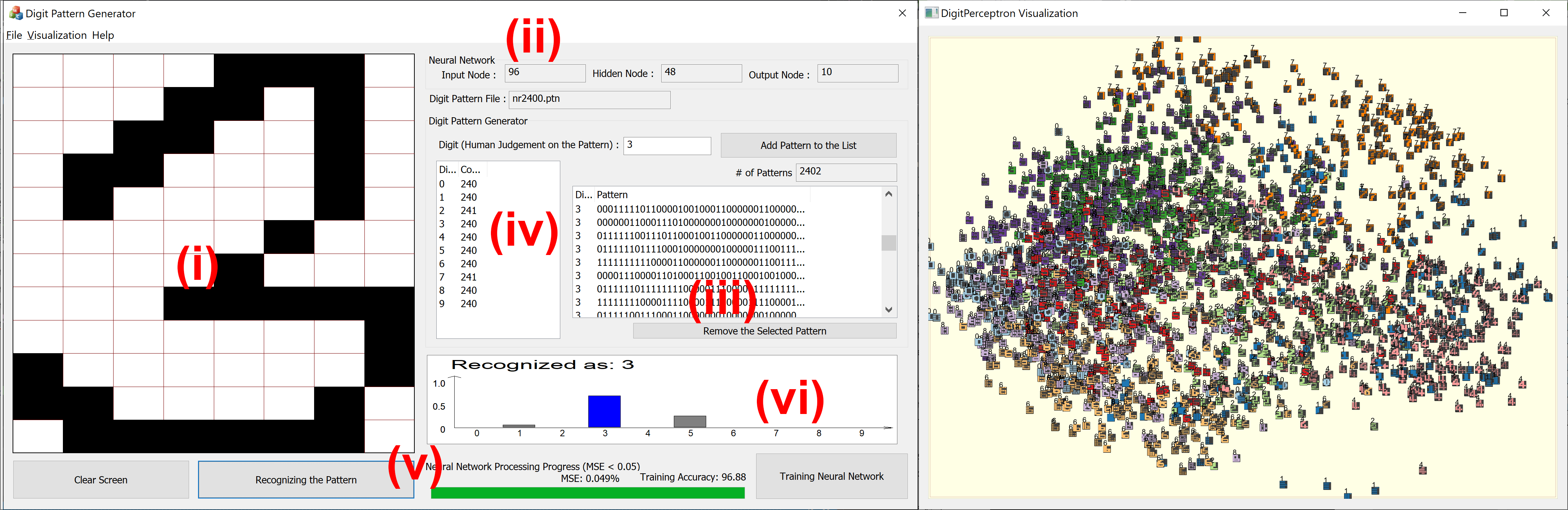}
        \caption{}
        \label{fig:system:a}
    \end{subfigure}
    \begin{subfigure}[b]{.34\linewidth}
        \includegraphics[height=2in, width=\linewidth]{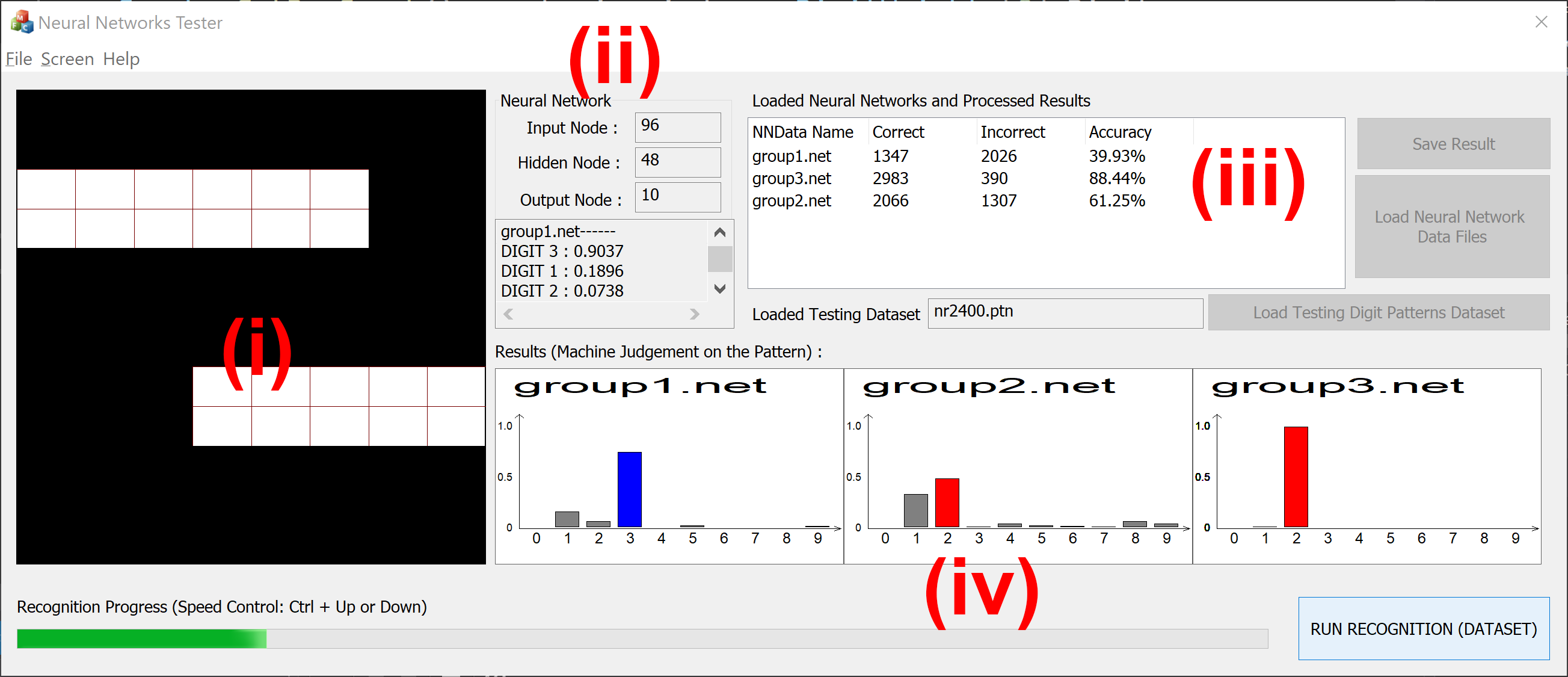}
        \caption{ }
        \label{fig:system:b}
    \end{subfigure}
 \caption{ Two systems are designed as (A) neural network trainer and (B) neural networks tester. The neural network trainer system consists of two layouts – digit pattern generator (left) and visual analyzer(right). The digit pattern generator supports the user in generating digit patterns and training neural networks. The visual analyzer represents user-generated digit patterns on a PCA projection space. The neural networks tester system evaluates multiple user-generated patterns with testing datasets.  }
 \label{fig:system}
\end{figure}

We hypothesized that supporting real-time interactive data generation, training, and recognition through NNs could increase users' understanding of the underlying idea of NNs. Based on this hypothesis, we have designed an active learning system (named Neural Network Trainer) to support the user in generating digit patterns and recognizing them with NNs. We also designed an additional system (called Neural Network Tester) to evaluate multiple user-generated patterns simultaneously. For supporting active learning in the system, integrating a graphical user interface (GUI) was considered to address the advancement of users' understanding of NNs through direct interactions with the system.
 
The Neural Network Trainer system consists of two layouts – Digit Pattern Generator (Figure \ref{fig:system:a}-left) and Visual Analyzer (Figure \ref{fig:system:a}-right). Digit Pattern Generator includes a pattern grid (Figure \ref{fig:system:a}-(i)) with multiple control panels (Figure \ref{fig:system:a}-(ii) $\sim$ \ref{fig:system:a}-(v)). The pattern grid allows the user to create digit patterns (i.e., 0 $\sim$ 9) by clicking each cell in the pattern grid. It has 12 × 8 grid cells representing a digit pattern. Each cell holds binary information as 1 or 0. It shows the size of NNs, including nodes in input, hidden, and output layers (Figure \ref{fig:system:a}-(ii)). Two list boxes have been added to keep all created digit patterns and the total number of patterns representing each digit in Figure \ref{fig:system:a}-(iii) and \ref{fig:system:a}-(iv), respectively. Real-time training and testing of NNs are handled with the control buttons (Figure \ref{fig:system:a}-(v)). The result of the recognized digit pattern with NN appears with probability distributions (Figure \ref{fig:system:a}-(vi)). Visual Analyzer represents user-generated digit patterns in a 2D display space by applying PCA computation. 

The Neural Network Tester system supports evaluating multiple user-generated NNs with various testing datasets. The primary purpose of having the system was to help the workshop participants understand the performances of their generated NNs in recognizing digits competitively with others. Figure \ref{fig:system:b} demonstrates the evaluation of three NNs created by three groups of users. Similar to the digit pattern generator, it has a pattern grid (Figure \ref{fig:system:b}-(i)) with multiple control panels (Figure \ref{fig:system:b}-(ii) $\sim$ \ref{fig:system:b}-(iv)). A list box (Figure \ref{fig:system:b}-(iii)) shows the loaded user-generated NNs. With a testing dataset, it evaluates the NNs showing overall accuracies (Figure \ref{fig:system:b}-(iii)) and probability estimation (Figure \ref{fig:system:b}-(iv)). The probability estimation indicates how each pattern is recognized with each NN. If a digit is recognized correctly, a reddish bar graph is represented. If not, a bluish bar graph is displayed to denote incorrect recognition. 

\section{Design of Neural Networks}
\label{sec:Design of Neural Networks}

To support real-time digit pattern generation and recognition, a three-layered NN based on the backpropagation method \cite{rojas96neural} was used. It feeds error rates back to NNs to optimize weights with optimal values. The input layer has 96 nodes to be matched to the cells of each digit pattern. The output layer has 10 nodes to represent digits 0 through 9. Although one or more hidden layers are often utilized in designing NNs, we have used one hidden layer consisting of 48 nodes in the system for speedy computation. For performance optimization, a gradient descent method was used because it could allow a parameter update of the weights. The sum of the squared error (SSE) was applied as the gradient of loss function L to determine the difference between the predicted ($\hat{y}_i$) and actual inputs ($y_i$) by: 

\begin{equation} \label{eq1}
L = - \frac{1}{N} \sum_{o = 0}^{N} \sum_{j = 0}^{C}(y_{o,j} - \hat{y}_{o,j})^2
\end{equation}
where $N$ is the length of digit samples, $C$ is the number of classes, and $y_{o,j}$ is an observation $y_o$ with a class $j$. 

To run NNs, momentum ($\gamma$) and learning rate ($\eta$) are defined to accelerate the training speed and accuracy of NNs. Momentum is a method that expedites the gradient descent by increasing the step size toward global minima. It is critical to find an optimal momentum value because a too-large value may skip global minima, or a too-small value may face local minimum issues. The learning rate controls how quickly a model adapts to the problem of training digit patterns. However, similar to tuning the momentum value, using an optimal learning rate is critical because it impacts the speed of the convergence to a solution and whether we can reach global optima. $\Delta W_{ij}$ and $\Delta W^{t-1}_{ij}$ represents weight changes in current and previous training iterations. They are given by:

\begin{equation} \label{iterations}
\Delta W_{ij}  = \gamma \Delta W^{t-1}_{ij} - \eta \frac{\sigma L}{\sigma W_{ij}},
\end{equation}
where $\frac{\sigma L}{\sigma W_{ij}}$ denotes the partial derivative of the loss function $L$ to decent update weights with learning rate $\eta$ with a multiplication of -1 to move towards global minima. The values of $\gamma$ and $\eta$ are determined based on empirical analysis for performance optimization in training data \cite{Bengio2012}. 

To activate nodes in the NNs, various activation functions are available, such as Sigmoid, ReLU (Rectified Linear Unit), Tanh, or hyperbolic tangent Activation Function.  ReLU is a broadly used activation function in convolutional neural networks (CNNs) or deep learning because it supports faster training \cite{Chou19}. However, it often causes a dying ReLU Problem \cite{Lu20} that decreases the ability of training data due to negative values becoming zero. Thus, the Sigmoid function, $ \sigma = \frac{1}{1 + \text{exp}^{-x}} $, is used in our system. It transforms the weighted sum of nodes to represent the probability of a value $x$ that belongs to a certain class. Although the Sigmoid activation function requires more computation than ReLU, it supports well for training a NN model in our designed system because it consists of a single hidden layer NN. To train NNs, termination condition  $\epsilon$ is defined as $\epsilon < 0.05$, which reduces the cost function L to become below the threshold. OpenMP API \cite{chandra2001parallel} is used to speed up the computation of NNs using multi-processors (i.e., multi-core processors).

\subsection{Pattern Generation}
\label{sec:Pattern Generation}

\begin{figure}[H]
    \centering
        \includegraphics[width=.5\linewidth]{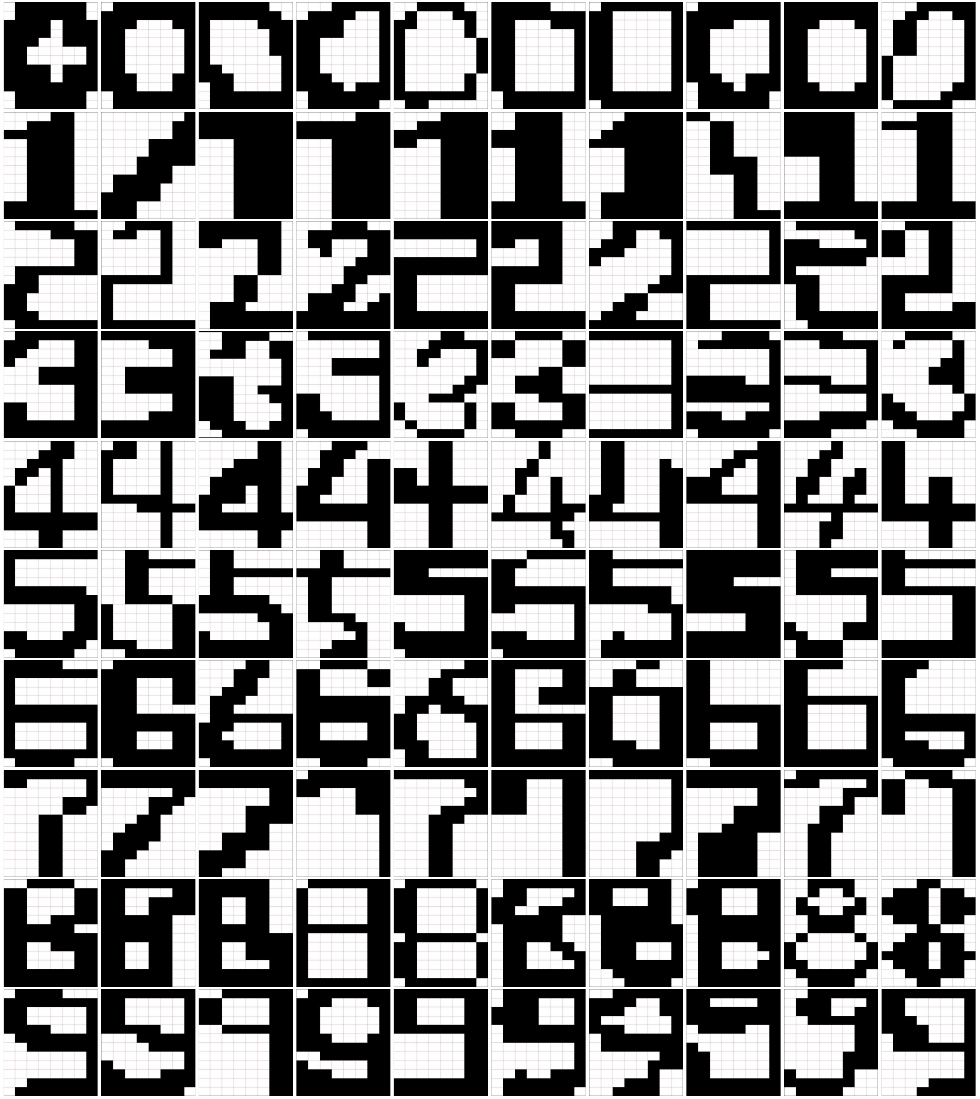}
\caption{Examples of user-generated digit patterns from the 2400-digit pattern dataset. Each pattern is created using the clickable pattern grid in the Neural Network Trainer system. }
\label{fig:mnist}
\end{figure}

To generate digit patterns, the user can enable or disable each cell using a computer mouse or a touch monitor screen (if available).  As mentioned earlier, the initial cell region has $12 \times 8$ size that supports creating up to $2^{12 \times 8}$ possible digit patterns. However, the overall number of patterns will be less because preprocessing generates duplicates by making each pattern fit into the cell region boundary. The applied preprocessing consists of three steps: (1) Determining the boundary of each pattern to find an object-bounding box; (2) Moving the pattern to the top-left corner; and (3) Applying scaling to make it fit the cell region. For scaling, a Nearest Neighbor Interpolation algorithm \cite{gonzalez2008digital} is used because it requires very litter calculations. Since each digit pattern has a binary color attribute (i.e., 0 or 1), each cell is marked if the interpolation satisfies the condition $I(x) > 0.5$. To help the user understand the internal preprocessing steps, intermediate outcomes become available only if a tracking option is enabled in the system.

The system supports saving user-generated digit patterns to a file and loading previously generated ones. To validate the effectiveness of the system, 2400-digit patterns are generated. Figure \ref{fig:mnist} shows samples of the 2400-digit patterns. When loading the previously generated digit patterns from a file, the system detects duplicated patterns and removes them if they exist.

\begin{figure}[H]
    \centering
    \begin{subfigure}[b]{0.2\linewidth}
        \includegraphics[width=\linewidth]{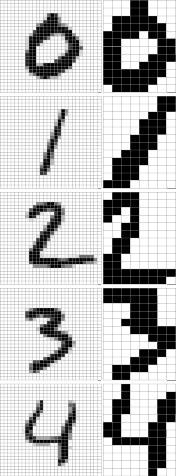}
    \end{subfigure} 
    \begin{subfigure}[b]{0.2\linewidth}
        \includegraphics[width=\linewidth]{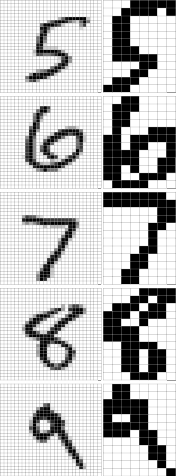}
    \end{subfigure} 
    \begin{subfigure}[b]{0.2\linewidth}
        \includegraphics[width=\linewidth]{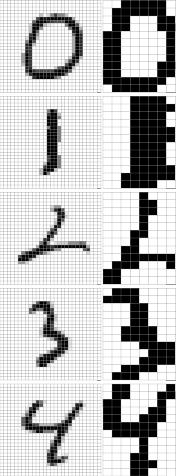}
    \end{subfigure} 
    \begin{subfigure}[b]{0.2\linewidth}
        \includegraphics[width=\linewidth]{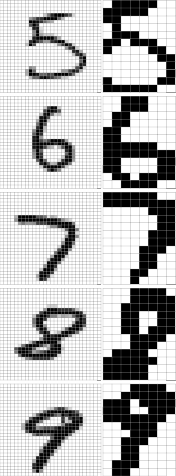}
    \end{subfigure} 
\caption{Conversion of the MNIST handwritten digits from the original images (28 $\times$  28 gray color) to two-tone colored images (12 $\times$ 8 binary color). The converted dataset is named TT-MNIST. }
\label{fig:tt-mnist}
\end{figure}

A handwritten digits dataset, MNIST \cite{deng2012mnist}, is also used to evaluate the system. It includes 70,000 grayscale images of handwritten digits with 60,000 training images and 10,000 testing images. Each digit sample is centered at a fixed-size image of $28 \times 28$ pixels. To make the images usable in our system, color conversion is applied to make them follow a binary color scheme using two-tone colors, black and white. Then, image size conversion is utilized to scale the image size down to $12 \times 8$ to make them fit into the clickable pattern grid in the designed system. Figure \ref{fig:tt-mnist} shows examples before and after applying the image conversion. Both gray color attribute conversion and image scaling are applied by using nearest neighbor interpolation with referencing neighbor color attributes. If the interpolated value meets the condition, i.e., $I(x) < \delta, 0 \le \delta \le 255$, the corresponding color attributes are changed to 0 when $I(x) < \delta$ and 1 when $I(x) \ge \delta$.We empirically determined an optimal value ($\delta = 85$) for converting MNIST digits to gray-colored images. For convenience, we call the user-generated 2400-digit samples and the converted two-tone colored MNIST images DS-2400 and TT-MNIST, respectively, in the rest of this paper. The two datasets are used to conduct a performance evaluation of the designed system. A detailed explanation about the conducted evaluation study is included in the evaluation section.

\subsection{Visual Representation}
\label{sec:Visualization}

\begin{figure}[H]
    \centering
        \includegraphics[width=0.9\linewidth]{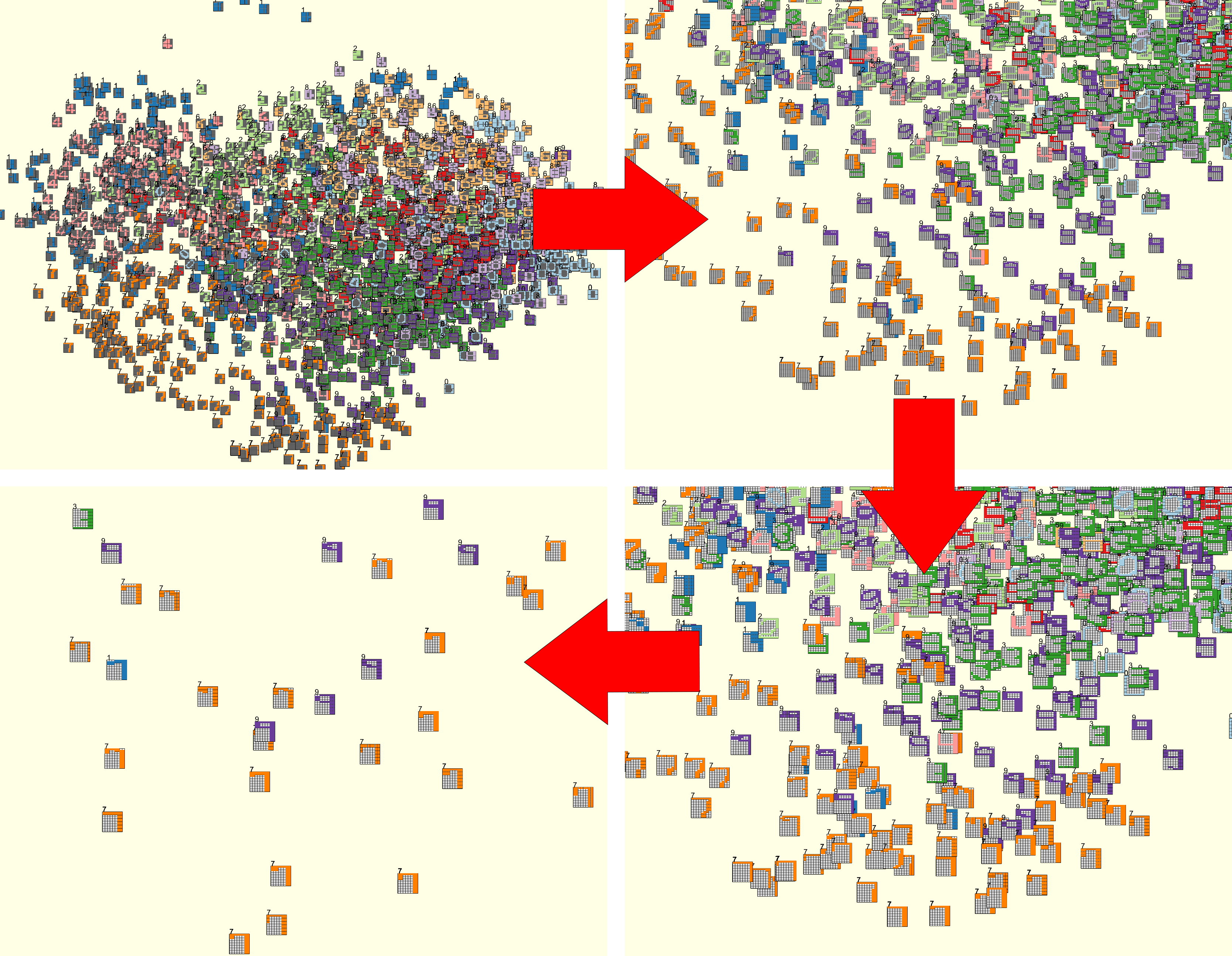}
\caption{Zooming and panning user interaction techniques are supported to navigate the PCA projection space. The user performed the zooming user interaction to see the region with the digit ``7'' patterns (located at the bottom of the space).}
\label{fig:visualization-zooming}
\end{figure}

\begin{figure}[H]
    \centering
    \begin{subfigure}[b]{.24\linewidth}
        \centering
        \includegraphics[height=1.4in, width=\linewidth]{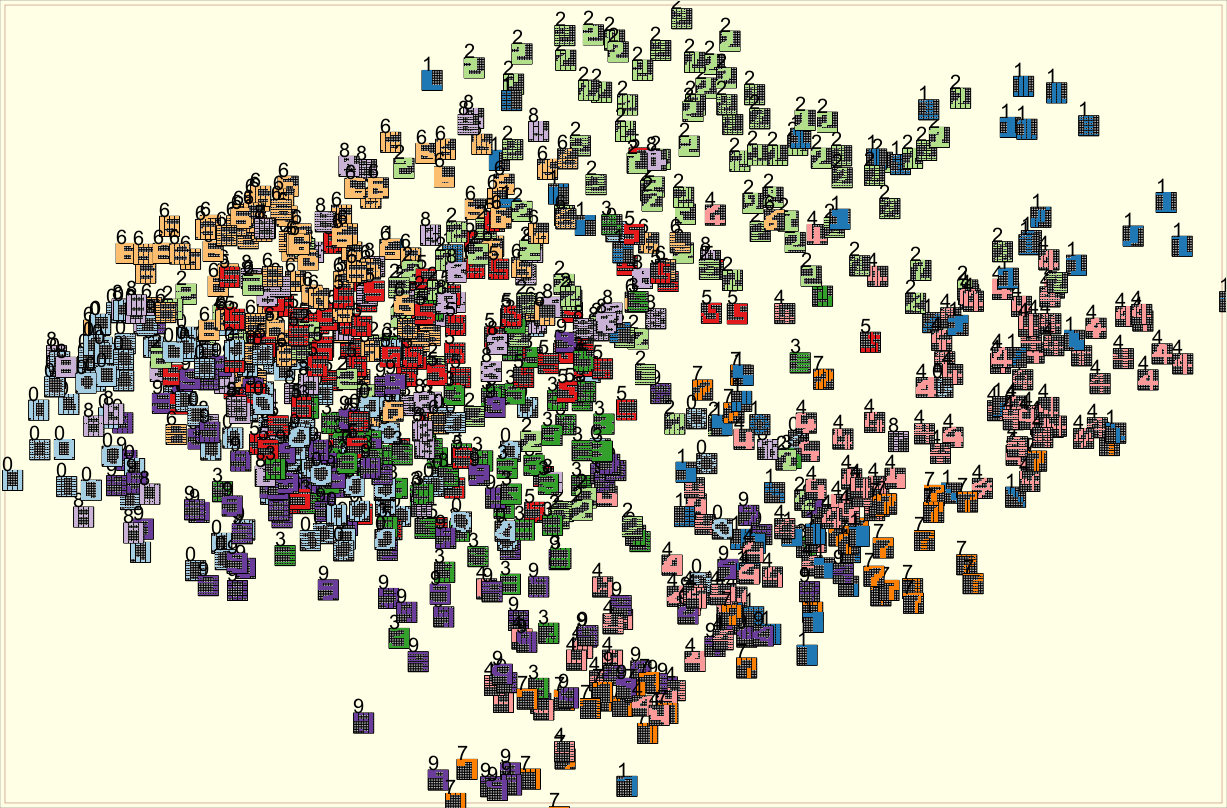}
        \caption{}
        \label{fig:pca:a}
    \end{subfigure}
    \begin{subfigure}[b]{.24\linewidth}
        \includegraphics[height=1.4in, width=\linewidth]{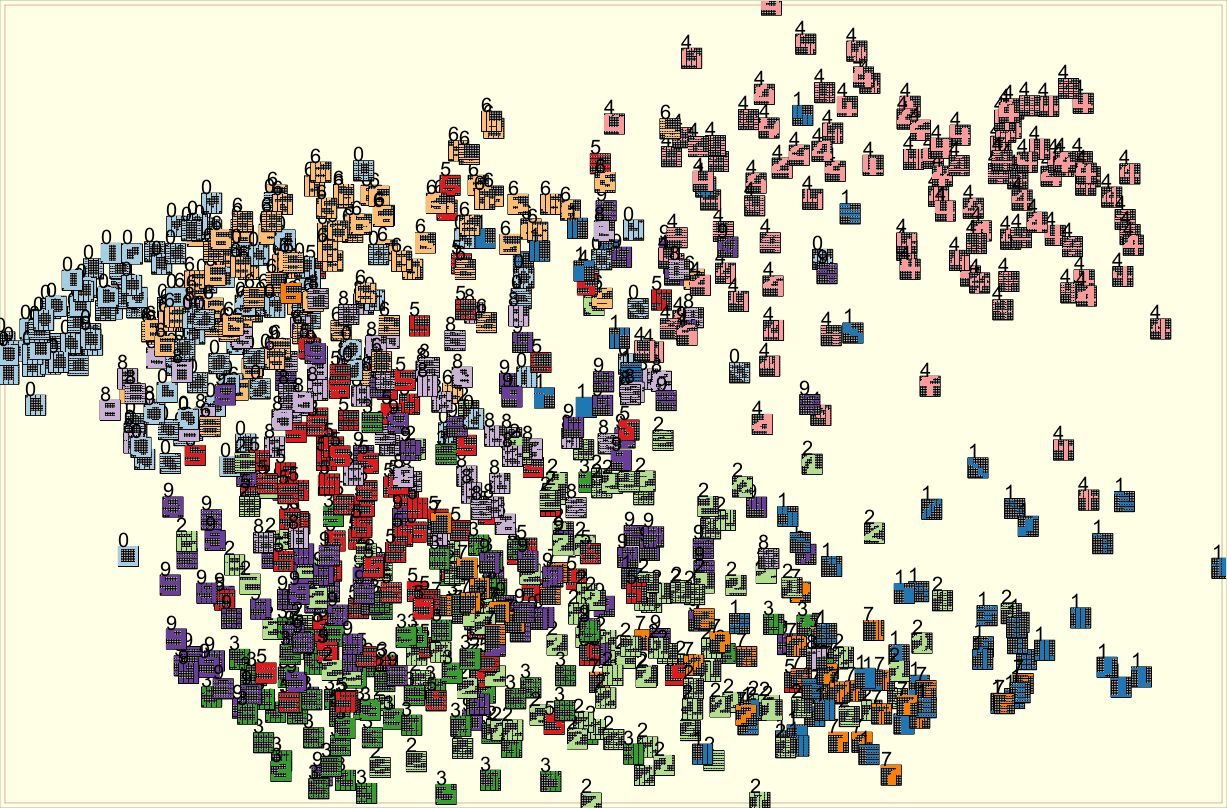}
        \caption{ }
        \label{fig:pca:b}
    \end{subfigure} 
    \begin{subfigure}[b]{.24\linewidth}
        \includegraphics[height=1.4in, width=\linewidth]{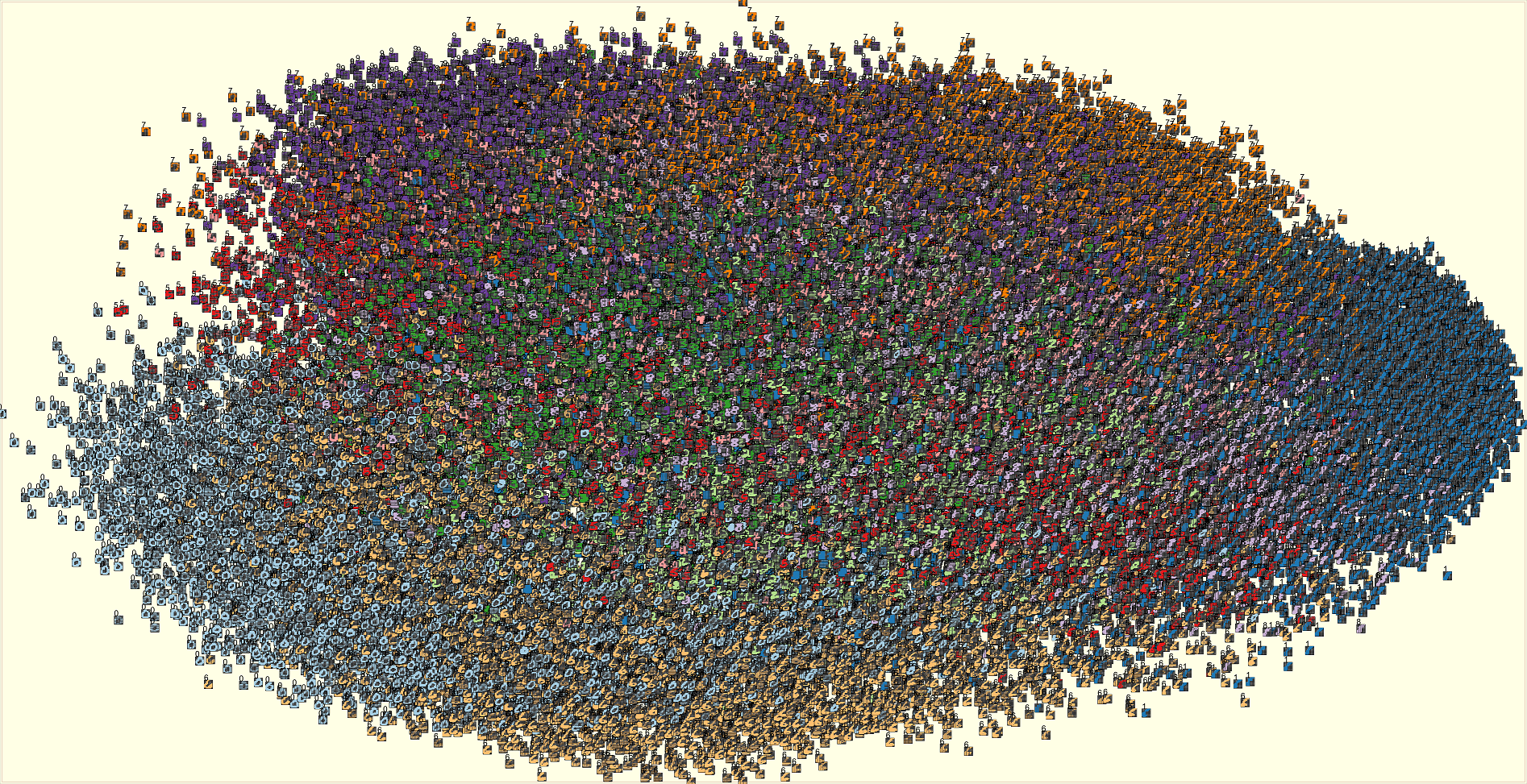}
        \caption{ }
        \label{fig:pca:c}
    \end{subfigure} 
    \begin{subfigure}[b]{.24\linewidth}
        \includegraphics[height=1.4in, width=\linewidth]{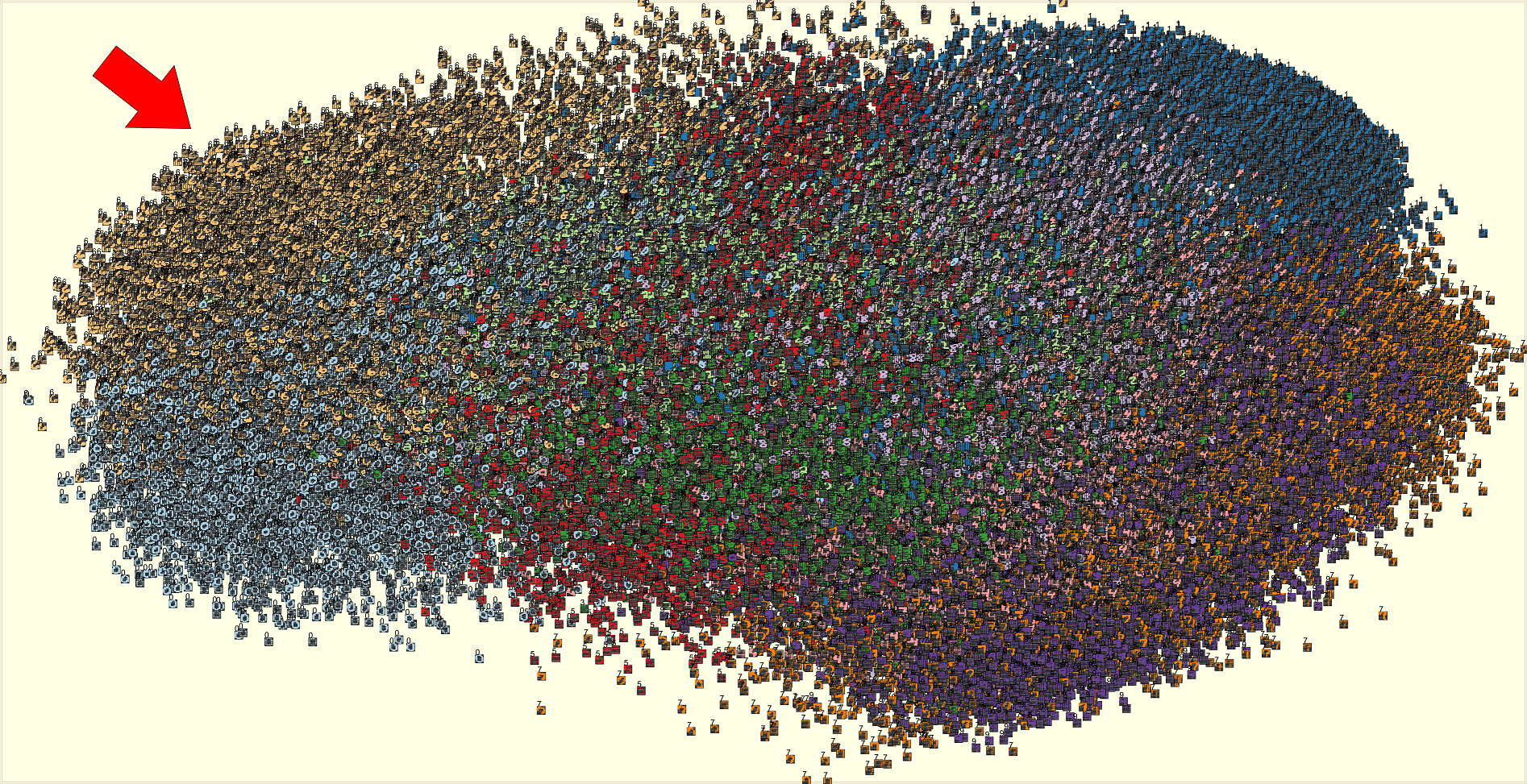}
        \caption{ }
        \label{fig:pca:d}
    \end{subfigure} 
    \caption{Visual representations of 940 digit patterns from the DS-2400 dataset in (a) and (b) and TT-MNIST dataset in (c) and (d). (a) and (c) show PCA projections using all digit patterns. (b) and (d) use hidden layer outputs as additional features. }\label{fig:pca}
\end{figure}

As mentioned above, a visualization representation is added to show all digit patterns to help users understand the difference between digits (see Figure \ref{fig:system:a}-right). Since each digit pattern consists of 96 cells, a dimension reduction technique, PCA is applied to project it in a PCA projection space (i.e., 2D display space). By default, the first and second principal components are used to display each digit pattern along the $x$- and $y$-axis.

The system supports basic navigational user interactions (i.e., zooming and panning) to help the user can navigate freely within the 2D display space to see the relationship among the digit patterns (see Figure \ref{fig:visualization-zooming}). It helps users understand the similarities between the digit patterns and the logic of recognizing them through NNs. Digit patterns maintaining similar cell outlines might appear nearby within the PCA projection space. As shown at the bottom of the visual representation, a digit 7's all patterns appear in a region. However, some digit 1's patterns appear near digit 7's patterns because they maintain similar markers, including vertical down-strokes with distinctive up-strokes on the top. Since digit 1's patterns do not always include the distinctive up-strokes, they appear in multiple regions. Similarly, different digit's patterns often appear in the same regions. The hidden layer outputs from trained NNs are used as additional features in PCA computation to create separable projections of digit patterns. Figure \ref{fig:pca} shows examples with the DS-2400 and TT-MNIST datasets. To help understand the usefulness of using the hidden layer outputs, randomly selected 940 digit patterns from the DS-2400 dataset are used to generate the projection with and without using the hidden layer outputs as additional features (see Figures \ref{fig:pca:a} and \ref{fig:pca:b}). It shows the benefit of using the hidden layout outputs by forming separated clusters among different digit patterns. With the TT-MNIST dataset, we observed a clear difference in Figures \ref{fig:pca:c} and \ref{fig:pca:d}. For instance, digit ``6'' patterns were observed in several locations in the PCA space (see Figure \ref{fig:pca:c}). But, with the integration of the hidden layout outputs, the digit patterns were positioned in the same region (see the arrow in Figure \ref{fig:pca:d}).

\section{Evaluation of the Interactive Learning System}
\label{sec:Evaluating the Effectiveness of the System}

\subsection{Interactive Learning}
\label{sec:Learing}

As discussed earlier, understanding how NNs work is not easy because of the complex nature of computing and updating its underlying structures continuously. The designed system may help users understand how it trains NNs and recognizes digit patterns. The system does not fully unveil its underlying structure of how the NN model changes its weights over time. However, we can conjecture that it supports interactive learning on NNs. More specifically, interactive learning manages three steps of the learning process: generating digit samples, training a NN model with the samples, and recognizing user-entered digits with the model. Generating digit patterns is essential to understand the effectiveness of NNs because it helps the user to identify how the NNs are trained to recognize digits. However, since it is not easy to create data, most studies have utilized existing datasets (e.g., MNIST, MS-COCO, ImageNet, Fashion-MNIST) to design new NN algorithms and evaluate their performances.

\begin{figure}[H]
    \centering
    \begin{subfigure}[b]{.48\linewidth}
        \includegraphics[height=1.2in, width=\linewidth]{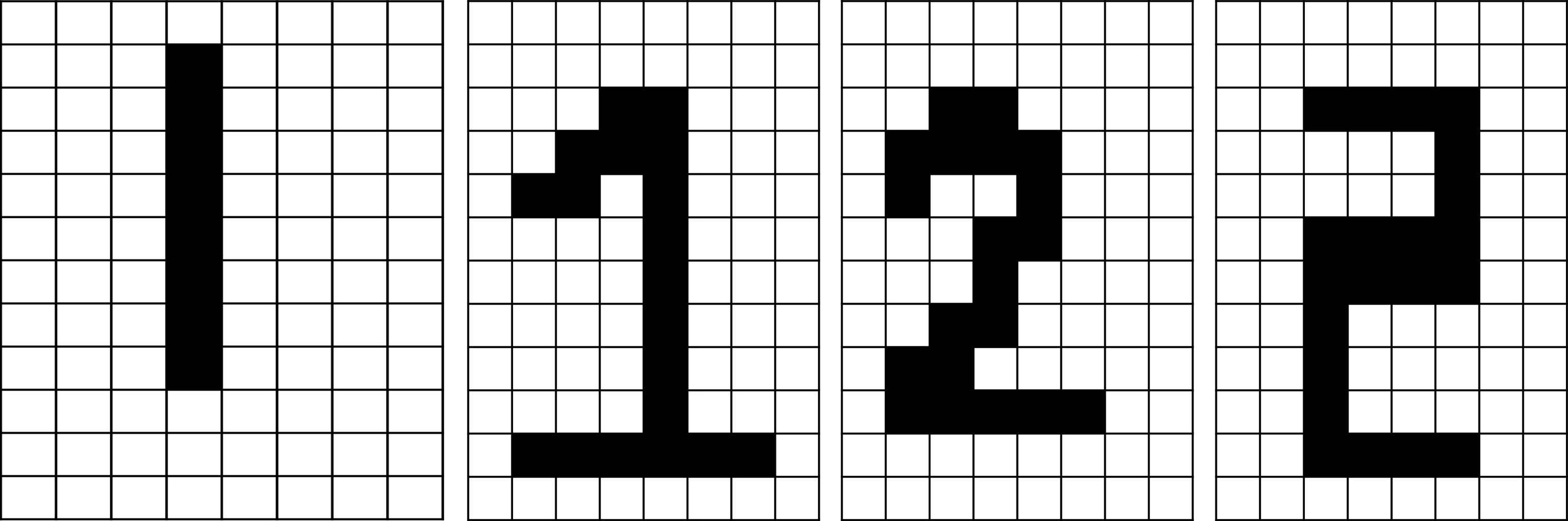}
        \caption{ }
        \label{fig:NNexample:a}
    \end{subfigure} 
    \begin{subfigure}[b]{.48\linewidth}
        \includegraphics[height=1.2in, width=\linewidth]{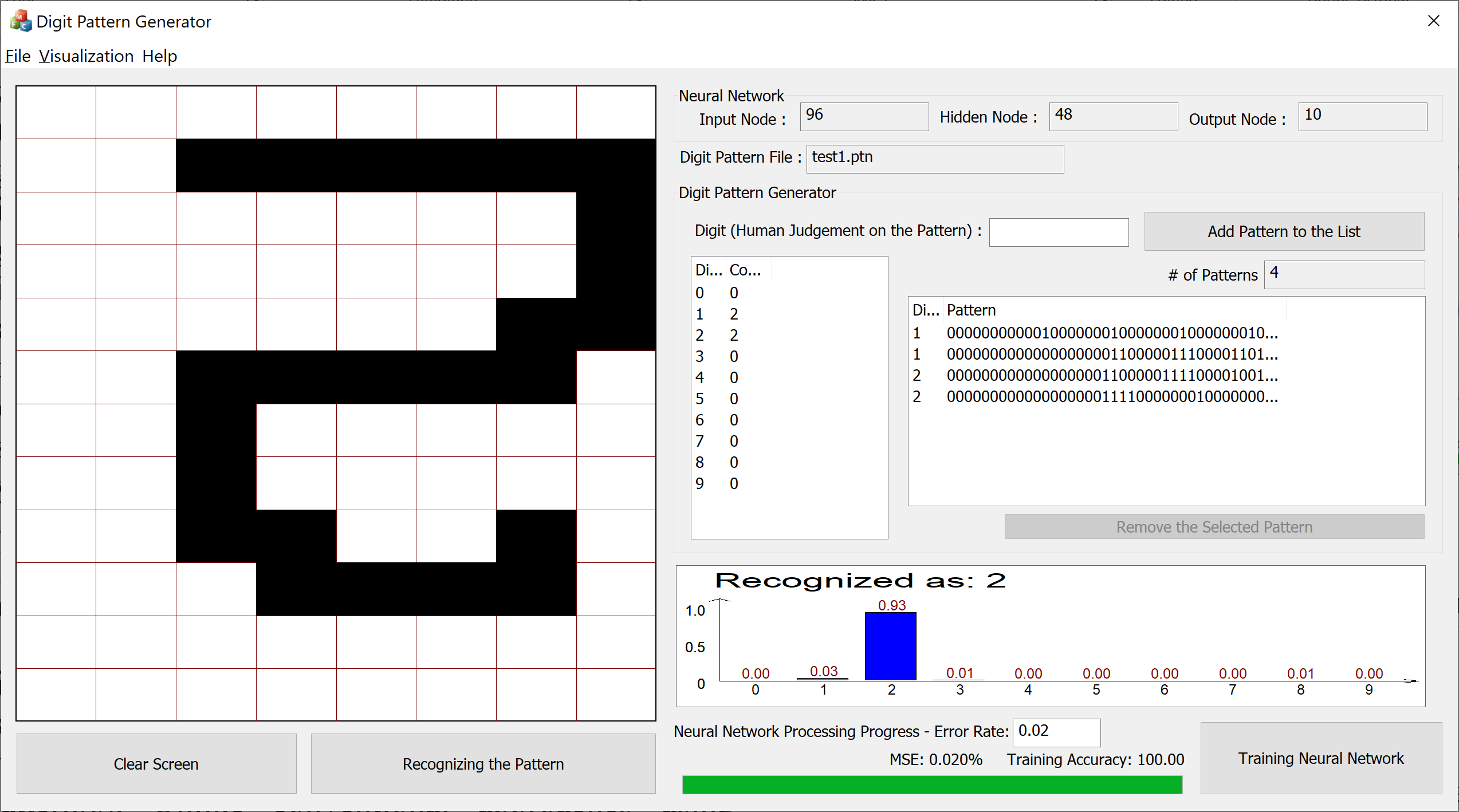}
        \caption{ }
        \label{fig:NNexample:b}
    \end{subfigure} 
\caption{An example of NN training with four-digit patterns (a) and recognizing a new digit with the trained NN (b).  }
\label{fig:NNexample}
\end{figure}

To support the user in creating digit patterns interactively, we used cell-based digit pattern generation to design simplified data digit samples using a computer mouse. The system allows training NNs whenever the user generates digit pattern(s). Unlike conventional approaches using numerous data samples, our system can train NNs with a small number of digit samples (e.g., $< 10$).  For instance, if a digit sample (denoting digit ``one'') is applied to train NNs, the same result (resulting digit ``one'') will be determined. Instead, if two distinctive digit samples (e.g., ``one'' and ``two'') are used to train NNs, the system correctly recognizes their differences. 
For example, if the user tries to recognize a new input pattern (similar to ``one'' or ``two'' digit patterns), the system correctly recognizes it as either one or two. Figure \ref{fig:NNexample} shows an example of recognizing a new digit pattern with four digit samples. Even though only four-digit patterns are used to train NNs, it correctly recognizes the new pattern with a high probability ($0.93$). The user can continuously add new digit patterns to improve the performance of recognizing digits interactively. This interactive digit pattern generation and recognition initiate active learning to help the user understand the logic behind NNs. For showing the probability, the probability distribution over all predicted classes is measured using a softmax function. It converts a vector of K values in the output layer to probability values. To show a normalized probability $[0,1]$ from the output value, the softmax function ($\sigma$) applies the exponential function. 

\begin{equation} \label{softmax}
\sigma(\vec{x}_i) = \frac{e^{x_{i}}}{\sum_{j=1}^K e^{x_{j}}}
\end{equation}
where $\vec{x}_i$ indicates the values in the NN output layer, $e^{x_i}$ and $e^{x_j}$ denote standard exponential function for output vector, respectively.

\subsection{Performance Evaluation}
\label{sec:performance testing}

\begin{figure}[H]
    \centering
    \begin{subfigure}[b]{0.24\linewidth}
        \includegraphics[width=\linewidth]{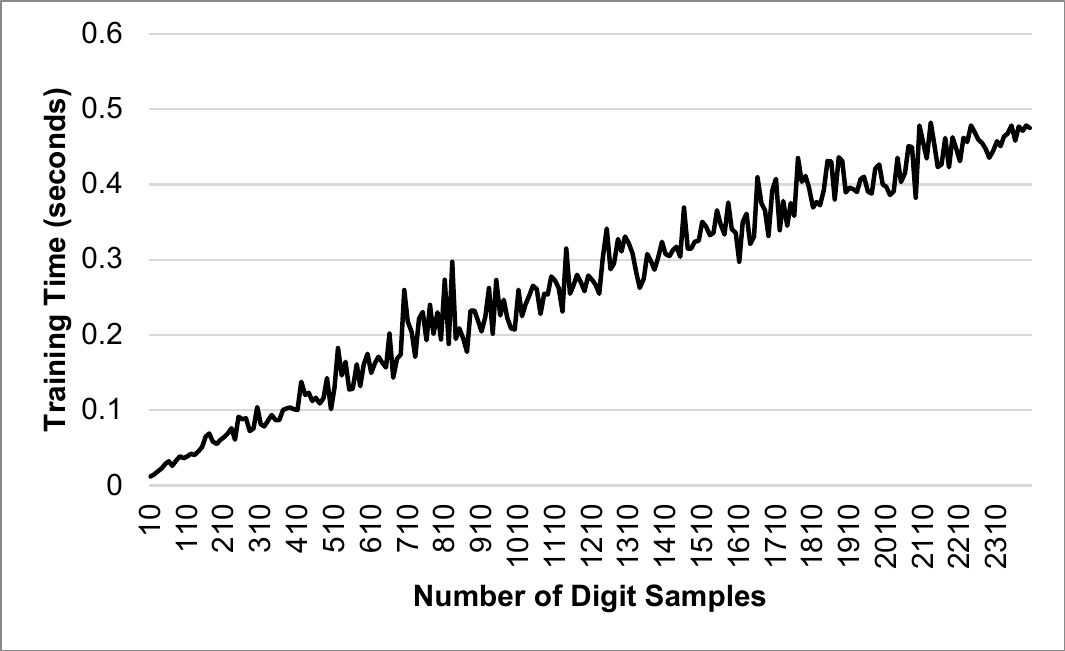}
        \caption{ }
        \label{fig:time_accuracy:a}
    \end{subfigure} 
    \begin{subfigure}[b]{0.24\linewidth}
        \includegraphics[width=\linewidth]{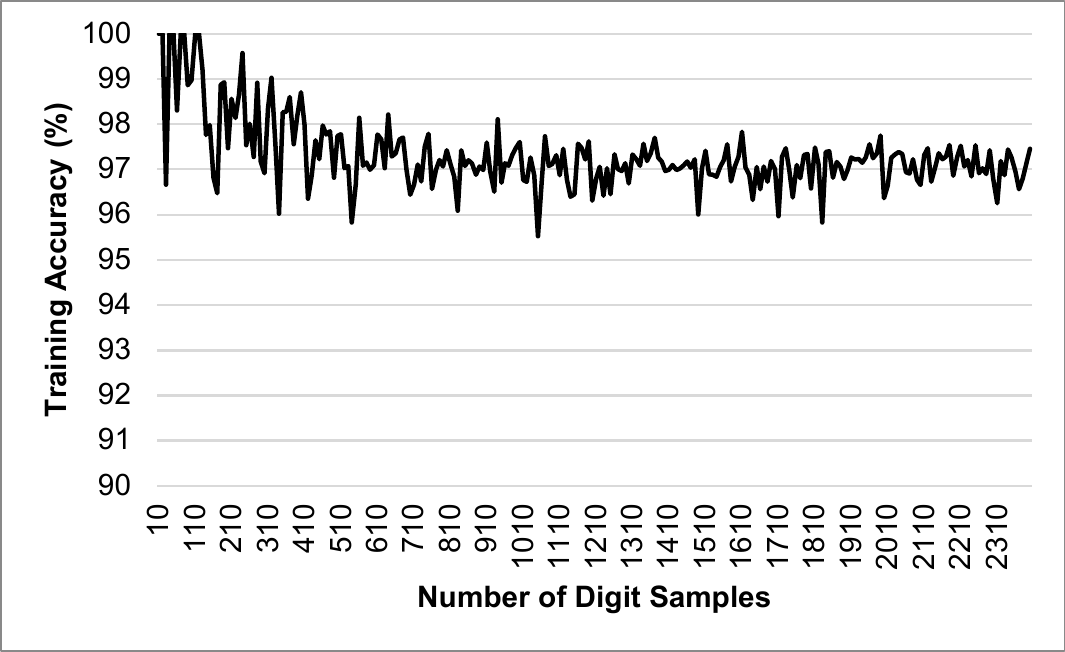}
        \caption{ }
        \label{fig:time_accuracy:b}
    \end{subfigure}  
    \begin{subfigure}[b]{0.24\linewidth}
        \includegraphics[width=\linewidth]{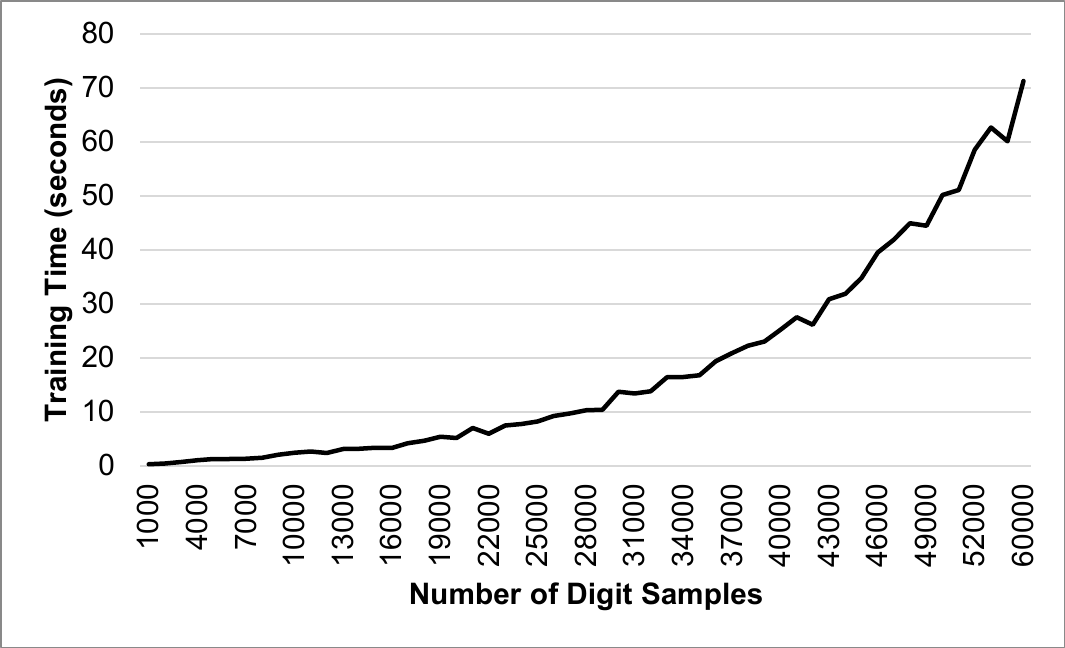}
        \caption{ }
        \label{fig:time_accuracy:c}
    \end{subfigure}  
    \begin{subfigure}[b]{0.24\linewidth}
        \includegraphics[width=\linewidth]{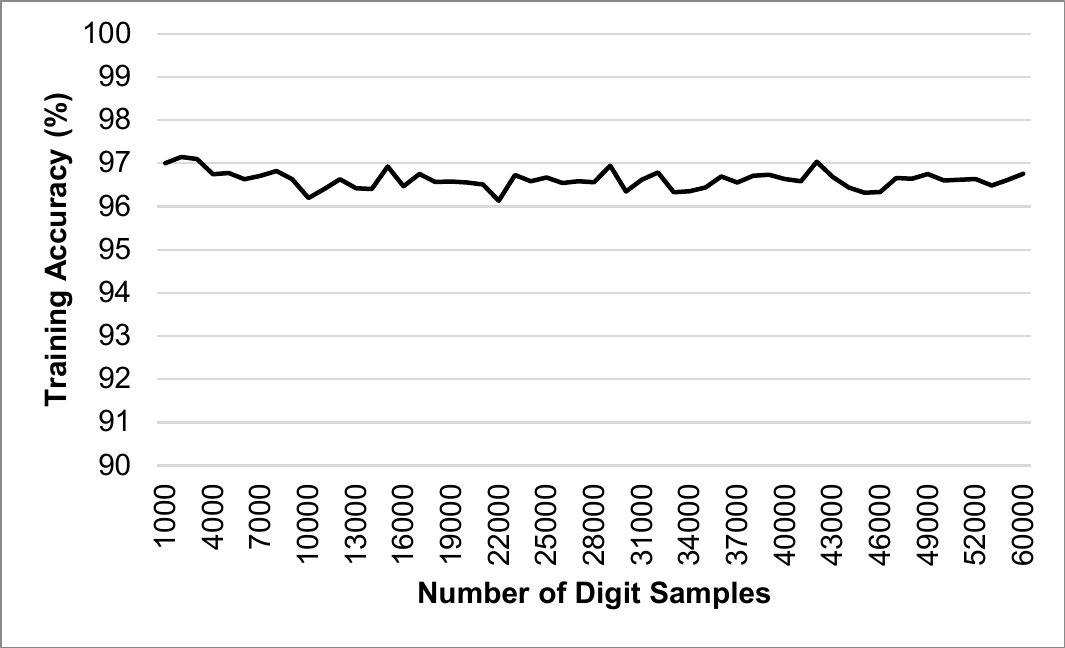}
        \caption{ }
        \label{fig:time_accuracy:d}
    \end{subfigure} 
\caption{Training time (seconds) and accuracy with different sizes of training data with (a) and (b) the DS-2400 dataset and (c) and (d) the TT-MNIST dataset. The $x-$axis shows the training digit sample size. (a) and (c) represent training time (seconds) and (b) and (d) show training accuracy.
}
\label{fig:time_accuracy}
\end{figure}

To support interactive learning, it is vital to maintain real-time training of NNs with the system while maintaining high accuracy. We performed an evaluation with Intel i9-9980HK Processor, 2.4 GHz, 8 cores. Figure \ref{fig:time_accuracy} shows that training time with the DS-2400 dataset was about $8.14 \pm 1.89$ seconds. With the DS-2400 dataset, the average training time was maintained to be less than 0.5 seconds. At the same time, the training accuracy was above 97\%. This indicates that the system supports users in performing real-time interactive analysis on digit recognition by training NNs. With the TT-MNIST dataset, we found that the training accuracy was maintained above 96\%. The training time gradually increases as the size of the samples grows. Approximately 70 seconds have been taken to train 60,000-digit patterns. Overall, the proposed system takes less than 2 seconds to train up to 10,000 digit patterns (training with MSE $< 0.05$, resulting in a training accuracy of $0.98$). Since the system supports real-time training on user-generated digit patterns, we can consider the system effectively helps the user understand how NNs work to effectively recognize the digits.

\section{Discussion}
\label{sec:Discussion}

As we mentioned above, the system is helpful for users to understand the logic behind NNs in recognizing digits. To understand how effective the system is in enhancing users' learning on NNs, we utilized the system during the workshop \footnote{
NSF funded 2022 Artificial Intelligence Awareness summer workshop: \url{https://csit.udc.edu/mudl/}
} for community college students. Most of the students do not have any knowledge or experience using NNs or related applications. They showed high interest in creating digit patterns and training NNs to recognize digits. Three groups were formed. They created patterns spending about 10 minutes  (see Table \ref{tbl:tab1}).  To understand the effectiveness of user-generated digit patterns, we tested the user-generated NNs using both the DS-2400 and TT-MNIST datasets. Although the testing accuracy was not high, we found that the trained NN by Group2 (using only 40 digit patterns) showed about 0.5 testing accuracy for the DS-2400 dataset.

\begin{table}[htbp]
\caption{Testing the NNs created by workshop participants with the two datasets (i.e., DS-2400 and TT-MNIST).}
\begin{center}
\begin{tabular}{|l|c|c|c|}
\hline
                                 & Group1 & Group2 & Group3 \\ \hline
Generated Patterns               & 29     & 40     & 13     \\ \hline
Testing Accuracy (with DS-2400)  & 0.29   & 0.50   & 0.32   \\ \hline
Testing Accuracy (with TT-MNIST) & 0.26   & 0.30   & 0.20   \\ \hline
\end{tabular}
\label{tbl:tab1}
\end{center}
\end{table}

The students commented that the system was highly interactive and useful for them to understand the underlying idea of NNs. They also reported the importance of utilizing both interactive pattern generation and visualization to upgrade their knowledge of NNs. Since the evaluation through workshop participants is purely informal user testing, it is essential to conduct formal user testing to examine the system's effectiveness in enhancing the user's understanding levels. Therefore, we plan to extend our study to conduct a formal user evaluation to validate the usefulness of the developed system.

\begin{figure}[H]
    \centering
    \begin{subfigure}[b]{.18\linewidth}
        \centering
        \includegraphics[width=\linewidth]{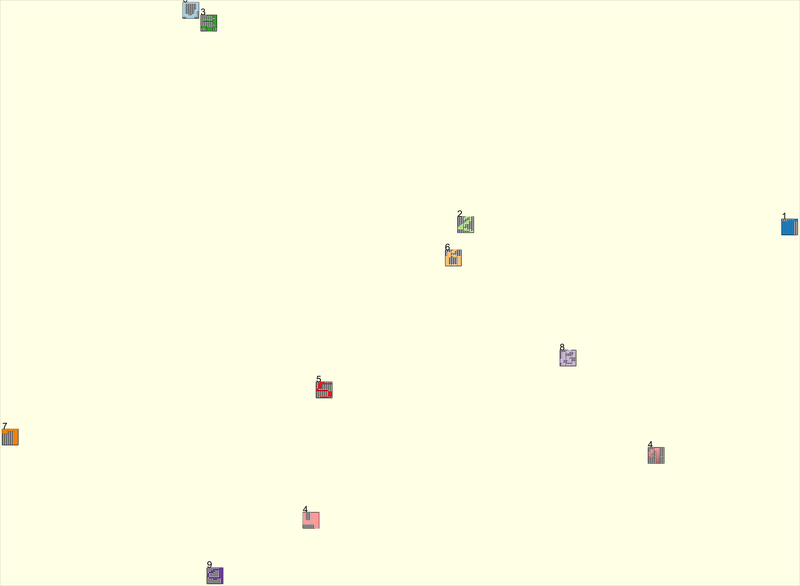}
        \caption{n = 10}
    \end{subfigure}
    \begin{subfigure}[b]{.18\linewidth}
        \includegraphics[width=\linewidth]{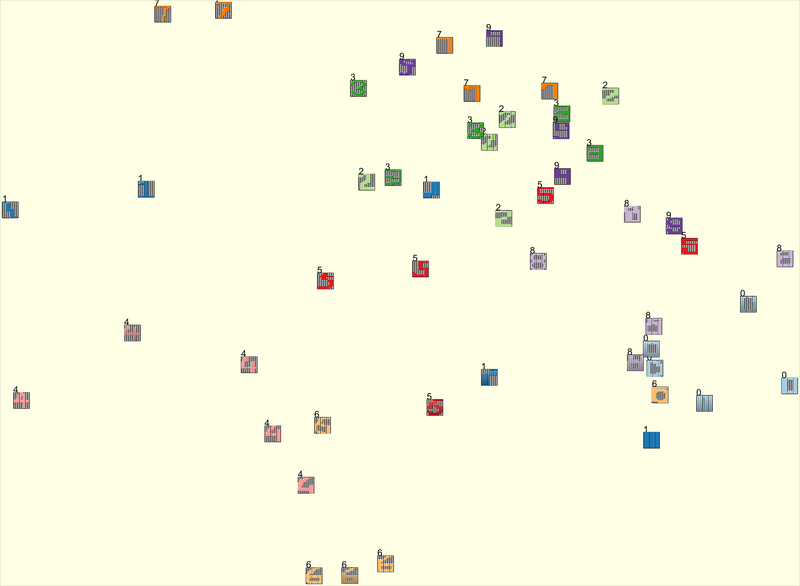}
        \caption{n = 50}
    \end{subfigure}
    \begin{subfigure}[b]{.18\linewidth}
        \includegraphics[width=\linewidth]{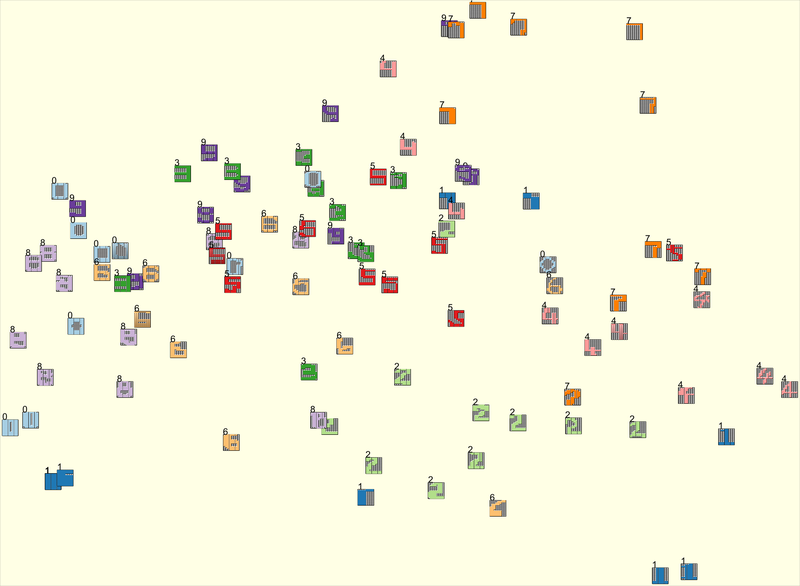}
        \caption{n = 100}
    \end{subfigure} 
    \begin{subfigure}[b]{.18\linewidth}
        \includegraphics[width=\linewidth]{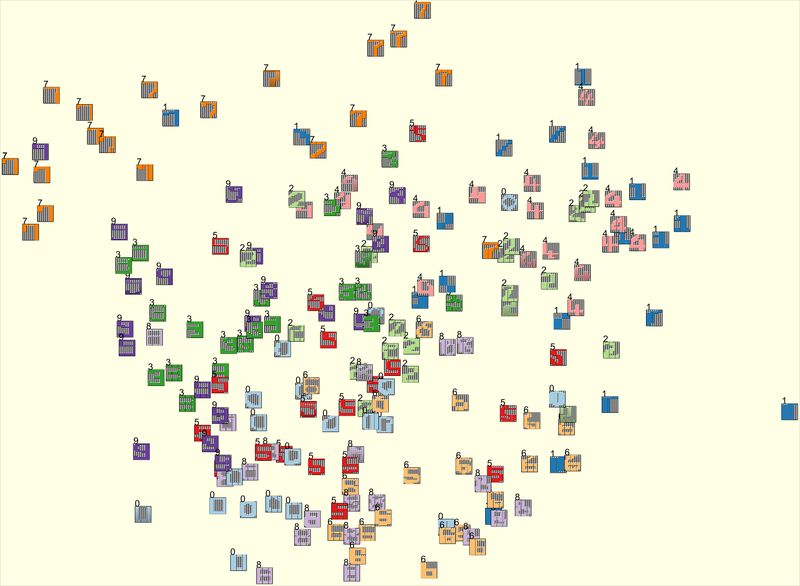}
        \caption{n = 200}
    \end{subfigure} 
    \begin{subfigure}[b]{.18\linewidth}
        \includegraphics[width=\linewidth]{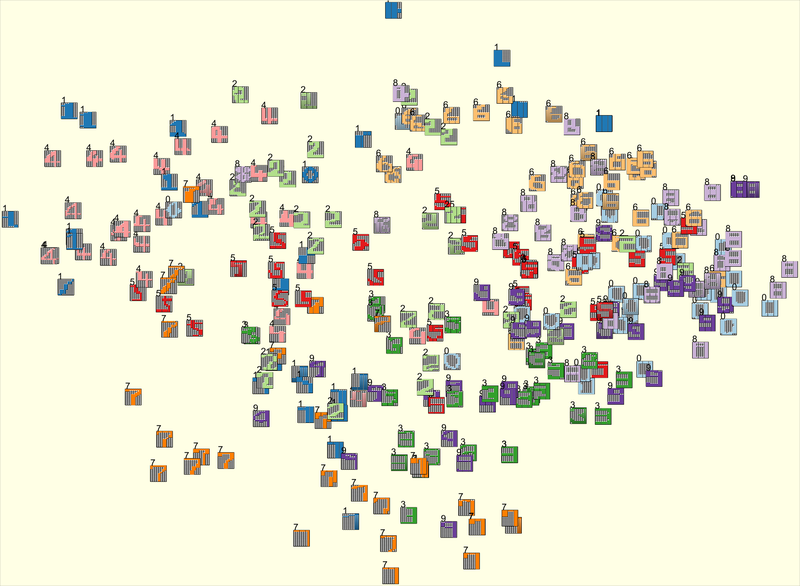}
        \caption{n = 300}
    \end{subfigure} 
    \newline
    \begin{subfigure}[b]{.18\linewidth}
        \centering
        \includegraphics[width=\linewidth]{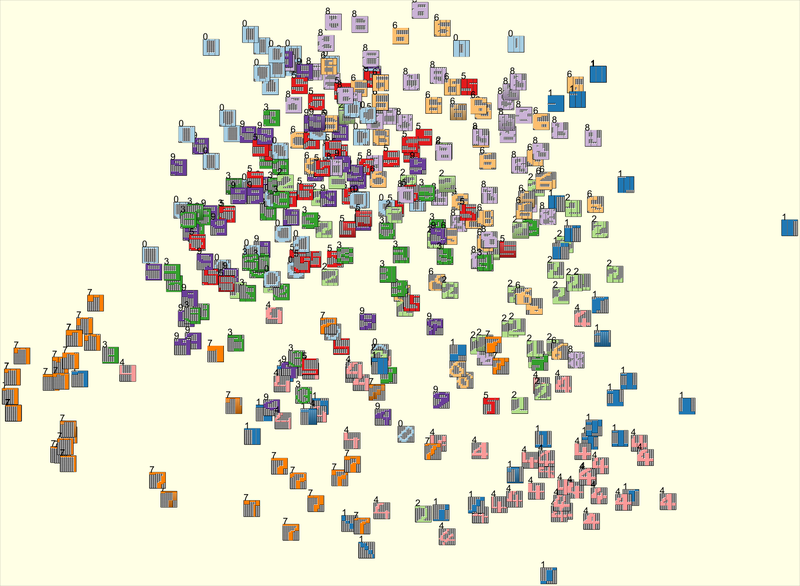}
        \caption{n = 400}
    \end{subfigure}
    \begin{subfigure}[b]{.18\linewidth}
        \includegraphics[width=\linewidth]{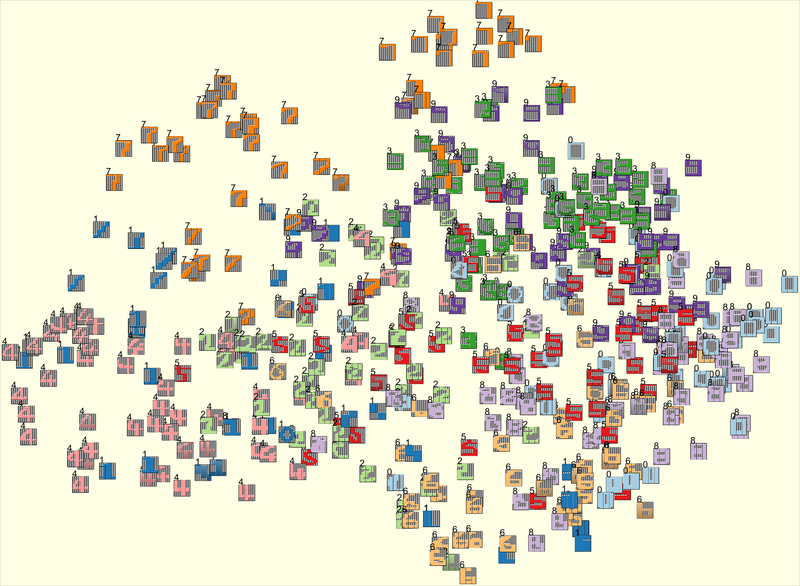}
        \caption{n = 500}
    \end{subfigure}
    \begin{subfigure}[b]{.18\linewidth}
        \includegraphics[width=\linewidth]{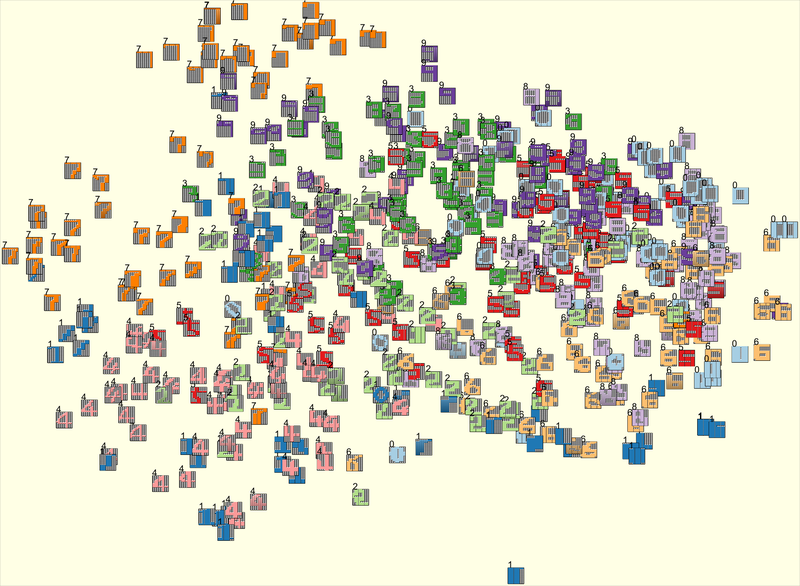}
        \caption{n = 600}
    \end{subfigure} 
    \begin{subfigure}[b]{.18\linewidth}
        \includegraphics[width=\linewidth]{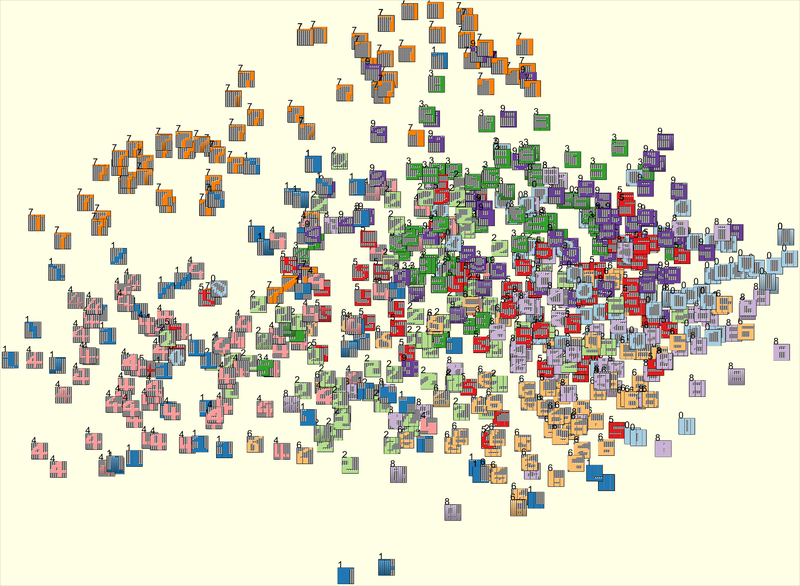}
        \caption{n = 700}
    \end{subfigure} 
    \begin{subfigure}[b]{.18\linewidth}
        \includegraphics[width=\linewidth]{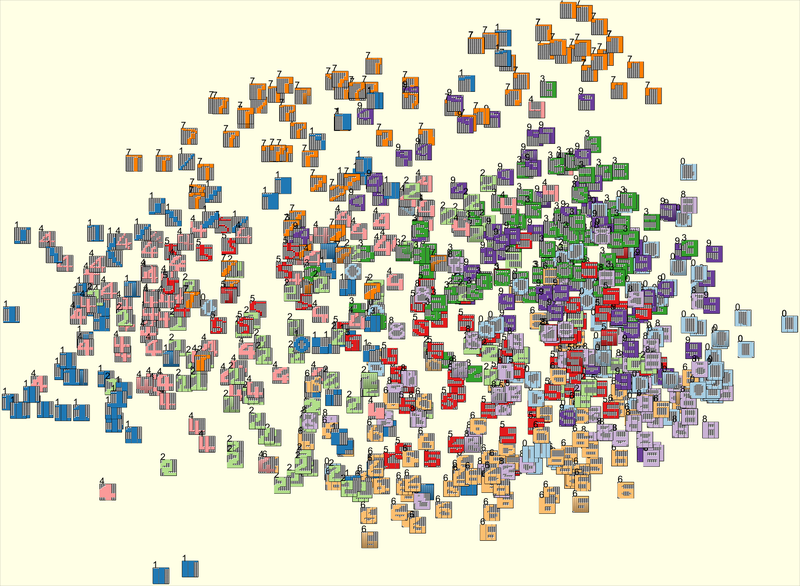}
        \caption{n = 800}
    \end{subfigure} 
    \newline
    \begin{subfigure}[b]{.18\linewidth}
        \centering
        \includegraphics[width=\linewidth]{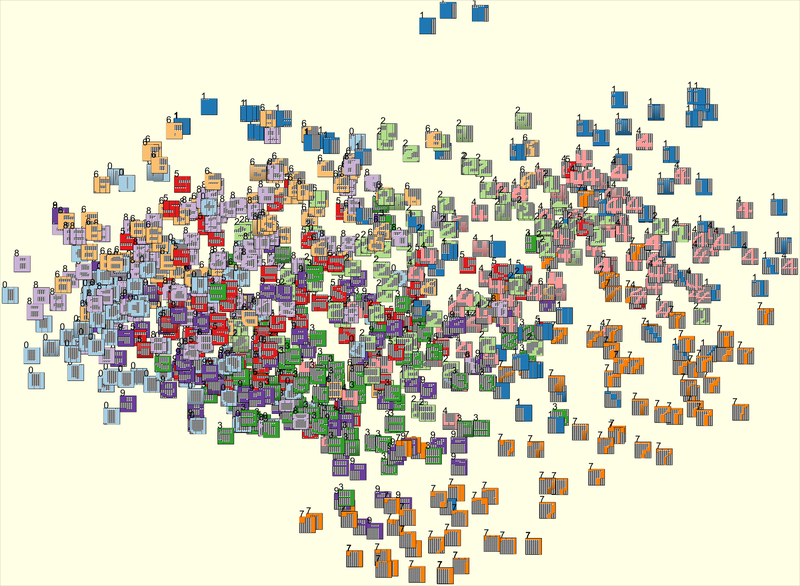}
        \caption{n = 900}
    \end{subfigure}
    \begin{subfigure}[b]{.18\linewidth}
        \includegraphics[width=\linewidth]{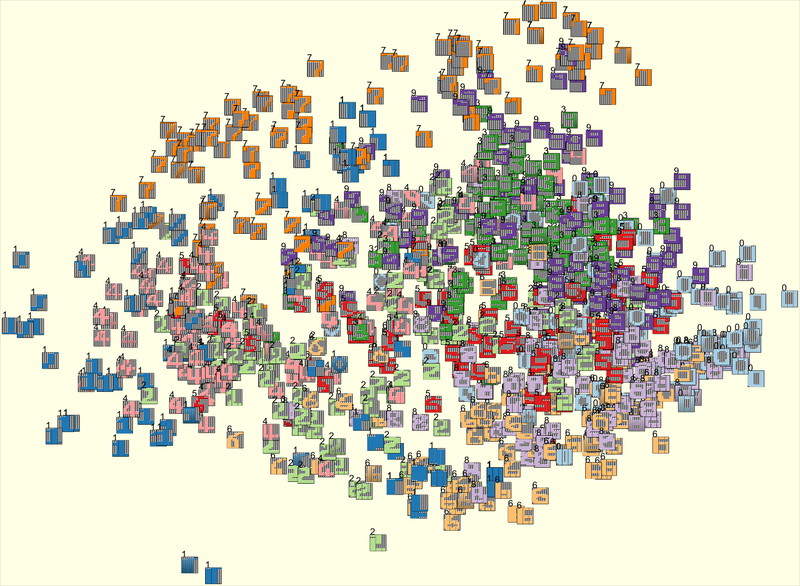}
        \caption{n = 1000}
    \end{subfigure}
    \begin{subfigure}[b]{.18\linewidth}
        \includegraphics[width=\linewidth]{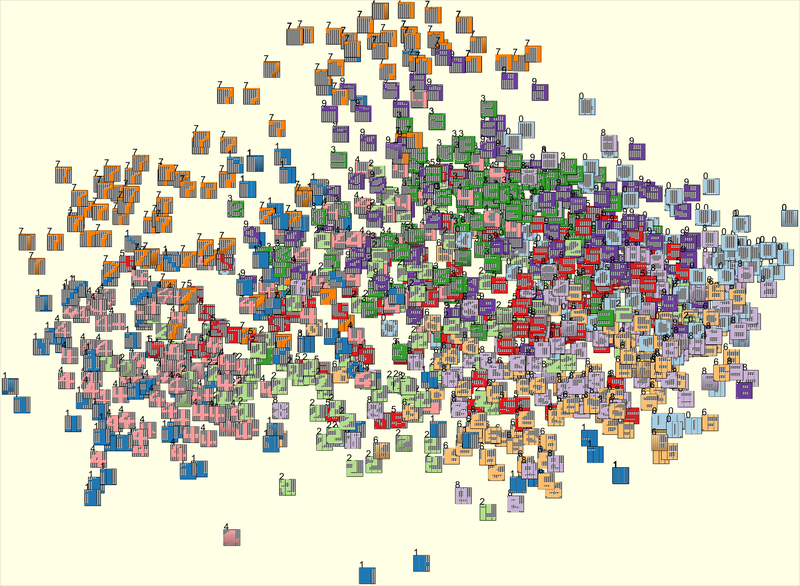}
        \caption{n = 1200}
    \end{subfigure} 
    \begin{subfigure}[b]{.18\linewidth}
        \includegraphics[width=\linewidth]{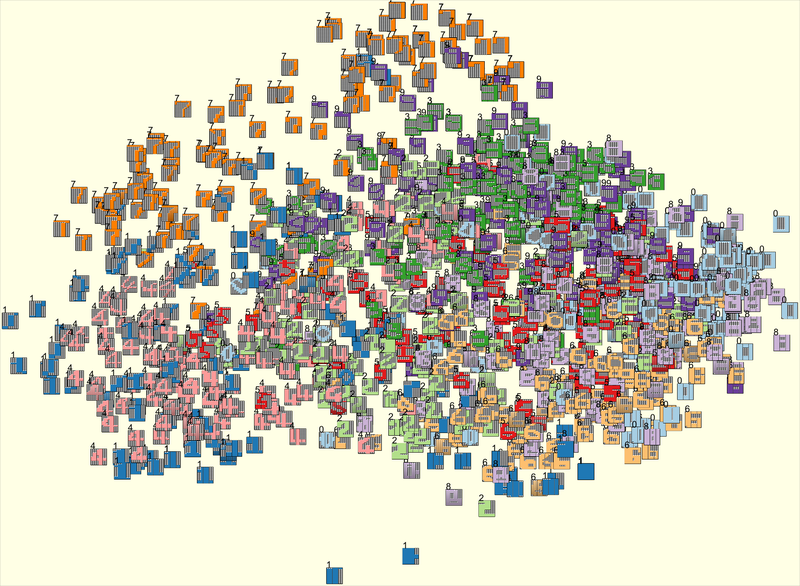}
        \caption{n = 1300}
    \end{subfigure} 
    \begin{subfigure}[b]{.18\linewidth}
        \includegraphics[width=\linewidth]{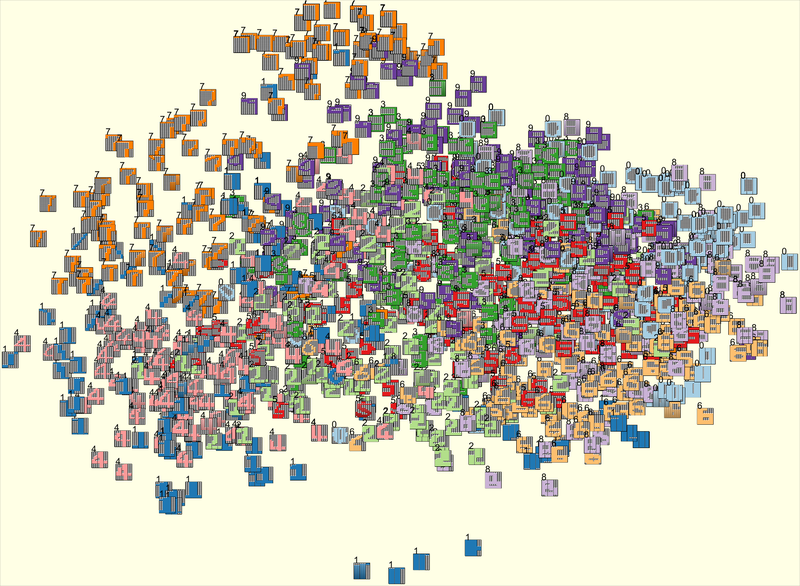}
        \caption{n = 1400}
    \end{subfigure} 
    \newline
    \begin{subfigure}[b]{.18\linewidth}
        \centering
        \includegraphics[width=\linewidth]{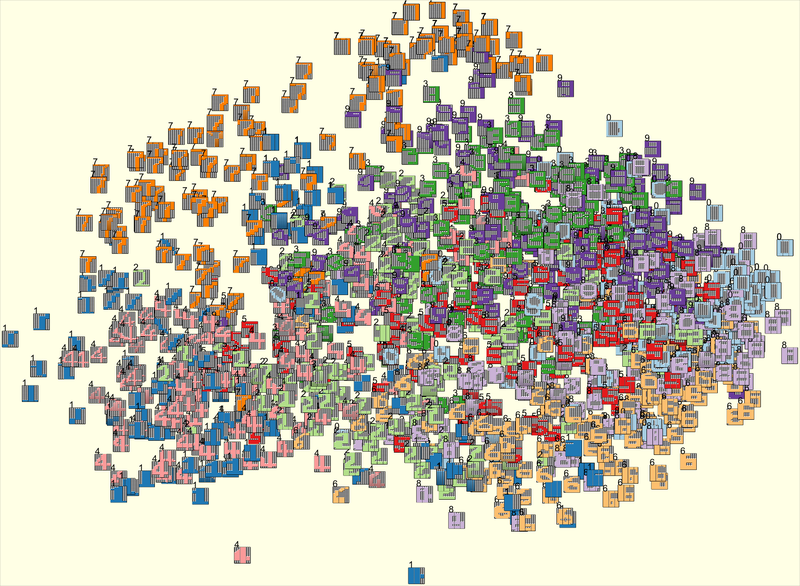}
        \caption{n = 1500}
    \end{subfigure}
    \begin{subfigure}[b]{.18\linewidth}
        \includegraphics[width=\linewidth]{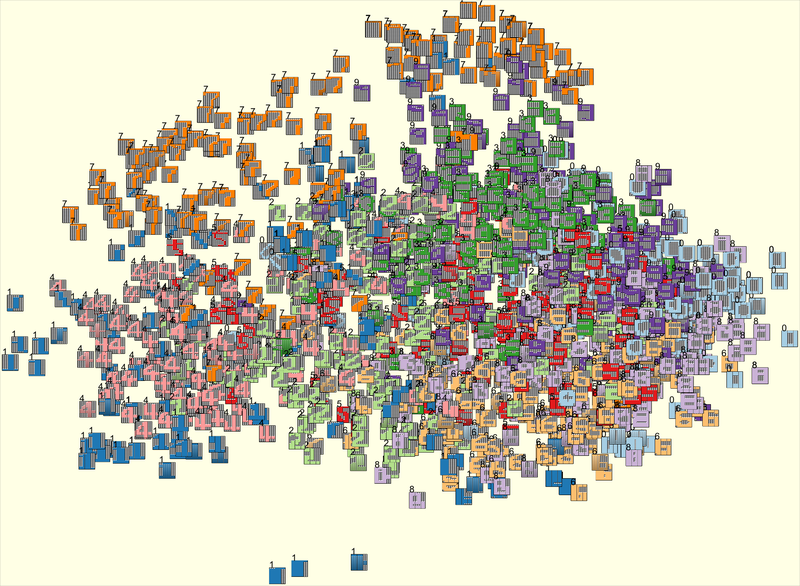}
        \caption{n = 1600}
    \end{subfigure}
    \begin{subfigure}[b]{.18\linewidth}
        \includegraphics[width=\linewidth]{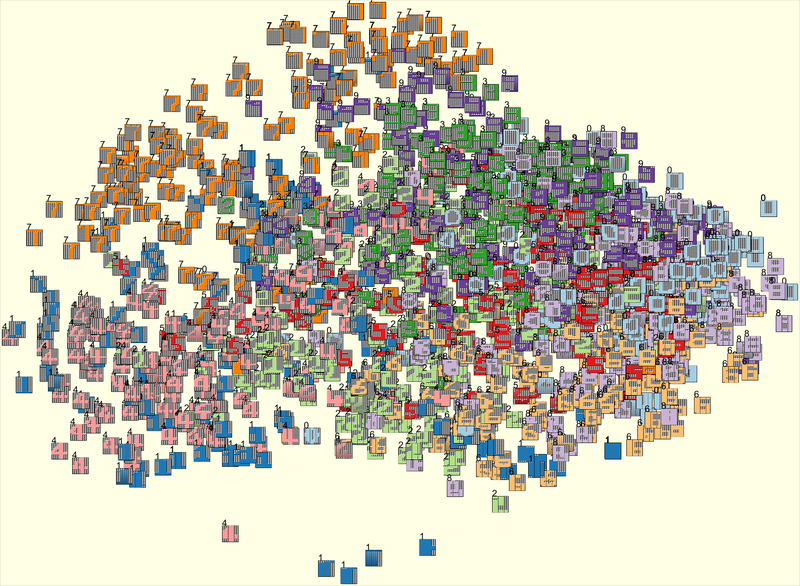}
        \caption{n = 1700}
    \end{subfigure} 
    \begin{subfigure}[b]{.18\linewidth}
        \includegraphics[width=\linewidth]{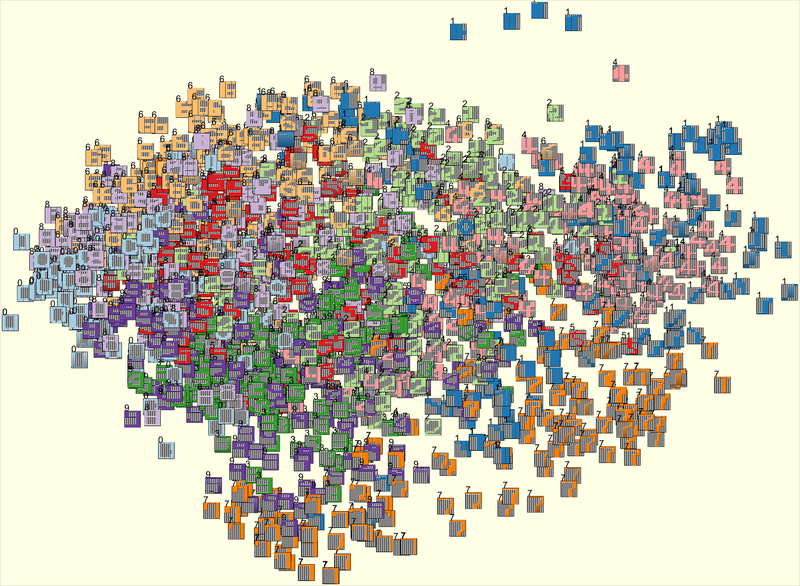}
        \caption{n = 1800}
    \end{subfigure} 
    \begin{subfigure}[b]{.18\linewidth}
        \includegraphics[width=\linewidth]{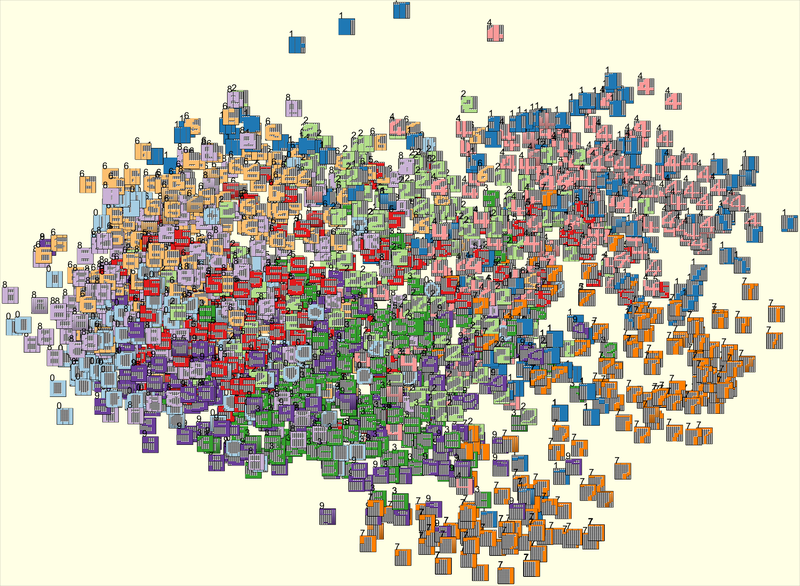}
        \caption{n = 1900}
    \end{subfigure} 
    \newline
    \begin{subfigure}[b]{.18\linewidth}
        \centering
        \includegraphics[width=\linewidth]{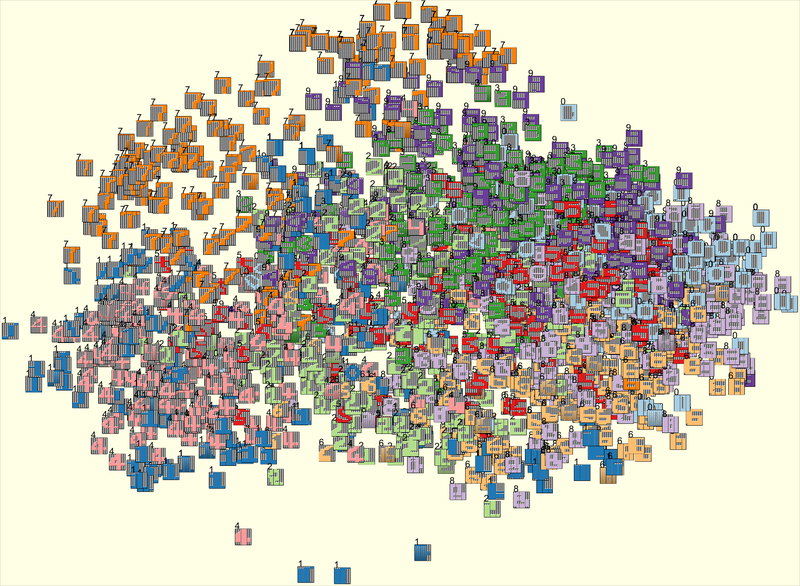}
        \caption{n = 2000}
    \end{subfigure}
    \begin{subfigure}[b]{.18\linewidth}
        \includegraphics[width=\linewidth]{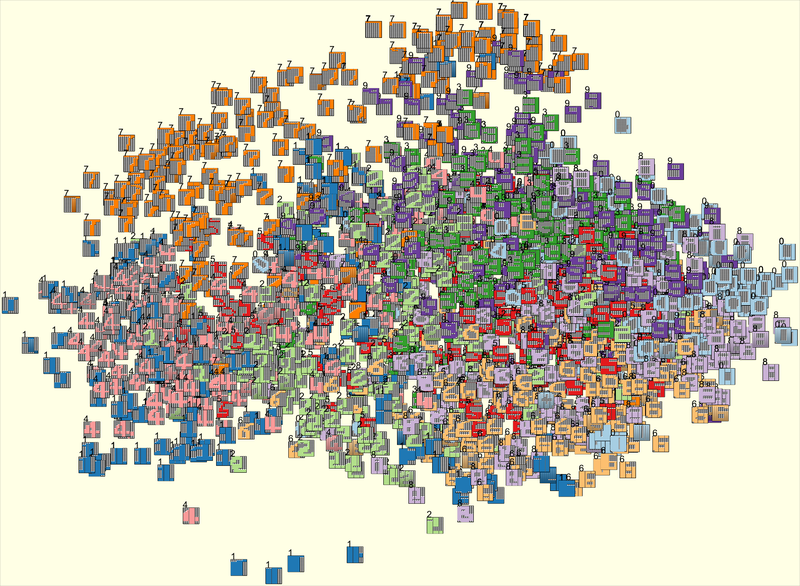}
        \caption{n = 2100}
    \end{subfigure}
    \begin{subfigure}[b]{.18\linewidth}
        \includegraphics[width=\linewidth]{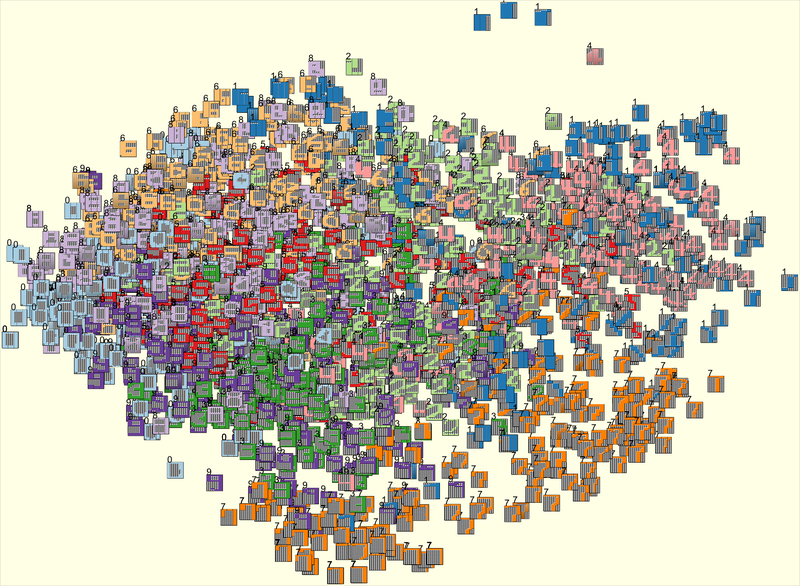}
        \caption{n = 2200}
    \end{subfigure} 
    \begin{subfigure}[b]{.18\linewidth}
        \includegraphics[width=\linewidth]{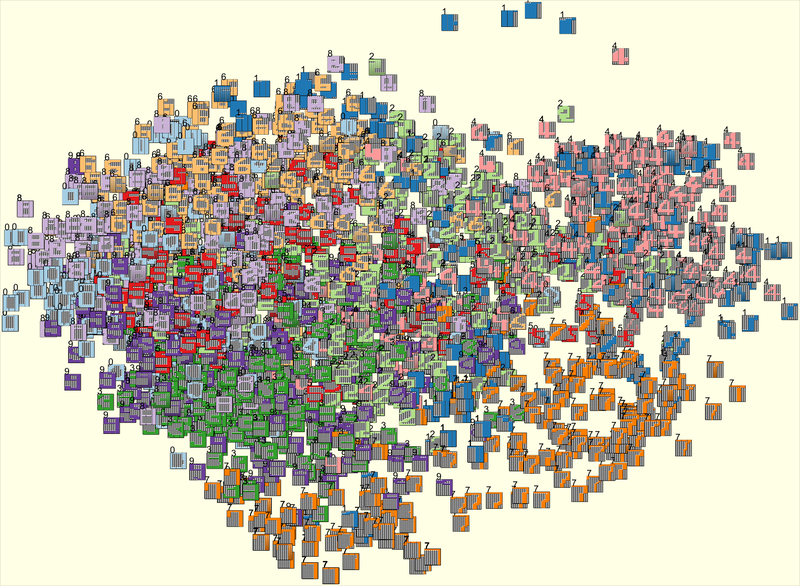}
        \caption{n = 2300}
    \end{subfigure} 
    \begin{subfigure}[b]{.18\linewidth}
        \includegraphics[width=\linewidth]{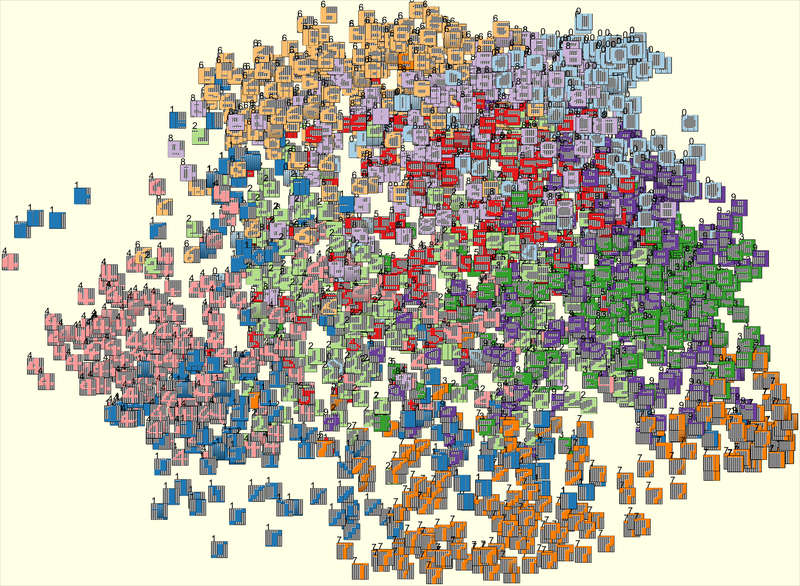}
        \caption{n = 2400}
    \end{subfigure} 
\caption{Visual representations of $n$ numbers of digit patterns from the DS-2400 dataset.
}
\label{fig:sign-flip}
\end{figure}

Although the system is useful for advancing users' understanding of NNs through interactive learning, it has a limitation of not showing the connection weights in NNs. Although representing the weight changes does not deliver additional information about understanding the internal changes of NNs, many researchers have emphasized the effectiveness of showing them \cite{DEPERLIOGLU2011392, you2019real}. Thus, designing an effective visual representation technique to show the connection weight changes in NNs is critical for advancing the users' knowledge of what information is effective in recognizing digits by NNs. It is important to note that PCA has an inherent ambiguity in the signs of resulting principal components. Thus, it generated multiple sign-flipped visual representations (see Figure \ref{fig:sign-flip}).

\section{Conclusion and Future Work}
\label{sec:Conclusion}

In this paper, we designed an interactive learning system to help users understand how neural networks perform digit recognition. The developed interactive learning system allows users to generate digit patterns and train them to recognize them in real time. To support real-time training and recognition, we introduced a simplified neural network with backpropagation to the system. Most importantly, we applied a visualization technique to show the difference among the digit patterns in a PCA projection space. In our experiments, we demonstrated the computational speed of training neural networks to evaluate the system's effectiveness. The key findings from this study are: (1) The developed interactive learning system took a short training time, which is critical for users to learn and understand NNs in real time; (2) The training accuracy was high (e.g., 96 $\sim$ 97\%) that validates the accuracy of the developed system as a tool to train NNs; and (3) Through our informal user testing based on the responses from community college students participated in the summer AI workshop, we received highly positive feedbacks although the number of the participants was fairly small (about ten participants).

For future work, we plan to conduct formal user testing to determine the system's effectiveness in terms of how much the user can understand the principle of neural networks. Since the system has been designed as a stand-alone application, a conversion of the system to a web-based application will be performed to make it become broadly available and accessible through a web browser. We will also extend the visual representation of digit samples with different dimensional reduction techniques, such as t-distributed stochastic neighbor embedding (t-SNE) \cite{Maaten08} or Uniform Manifold Approximation and Projection (UMAP) \cite{mcinnes2018umap}. The complete codes and a pre-compiled executable are available at \url{https://github.com/drjeong/DigitPerceptron}

\section*{Acknowledgments}
This material is based upon work supported by the National Science Foundation under Grant No. (2107449, 2107450, and 2107451).


\begin{thebibliography}{10}

\bibitem{miller2018artificial}
D~Douglas Miller and Eric~W Brown.
\newblock Artificial intelligence in medical practice: the question to the
  answer?
\newblock {\em The American journal of medicine}, 131(2):129--133, 2018.

\bibitem{du2016recent}
Benedict du~Boulay.
\newblock Recent meta-reviews and meta--analyses of aied systems.
\newblock {\em International Journal of Artificial Intelligence in Education},
  26(1):536--537, 2016.

\bibitem{karasozen2020earthquake}
Ezgi Karas{\"o}zen and B{\"u}lent Karas{\"o}zen.
\newblock Earthquake location methods.
\newblock {\em GEM-International Journal on Geomathematics}, 11(1):1--28, 2020.

\bibitem{duan2019artificial}
Yanqing Duan, John~S Edwards, and Yogesh~K Dwivedi.
\newblock Artificial intelligence for decision making in the era of big
  data--evolution, challenges and research agenda.
\newblock {\em International journal of information management}, 48:63--71,
  2019.

\bibitem{Popenici2017}
Stefan A.~D. Popenici and Sharon Kerr.
\newblock Exploring the impact of artificial intelligence on teaching and
  learning in higher education.
\newblock {\em Research and Practice in Technology Enhanced Learning},
  12(1):22, Nov 2017.

\bibitem{wang2020should}
Fei Wang, Rainu Kaushal, and Dhruv Khullar.
\newblock Should health care demand interpretable artificial intelligence or
  accept “black box” medicine?, 2020.

\bibitem{BARREDOARRIETA202082}
Alejandro {Barredo Arrieta}, Natalia Díaz-Rodríguez, Javier {Del Ser}, Adrien
  Bennetot, Siham Tabik, Alberto Barbado, Salvador Garcia, Sergio Gil-Lopez,
  Daniel Molina, Richard Benjamins, Raja Chatila, and Francisco Herrera.
\newblock Explainable artificial intelligence (xai): Concepts, taxonomies,
  opportunities and challenges toward responsible ai.
\newblock {\em Information Fusion}, 58:82--115, 2020.

\bibitem{HOLZINGER2022263}
Andreas Holzinger, Matthias Dehmer, Frank Emmert-Streib, Rita Cucchiara,
  Isabelle Augenstein, Javier~Del Ser, Wojciech Samek, Igor Jurisica, and
  Natalia Díaz-Rodríguez.
\newblock Information fusion as an integrative cross-cutting enabler to achieve
  robust, explainable, and trustworthy medical artificial intelligence.
\newblock {\em Information Fusion}, 79:263--278, 2022.

\bibitem{Haresamudram2023}
Kashyap Haresamudram, Stefan Larsson, and Fredrik Heintz.
\newblock Three levels of ai transparency.
\newblock {\em Computer}, 56(2):93--100, 2023.

\bibitem{Bathaee2018TheAI}
Yavar Bathaee.
\newblock The artificial intelligence black box and the failure of intent and
  causation.
\newblock {\em Harvard Journal of Law \& Technology}, 31:889, 2018.

\bibitem{roweis1997algorithms}
Sam Roweis.
\newblock Em algorithms for pca and spca.
\newblock {\em Advances in neural information processing systems}, 10, 1997.

\bibitem{Jeong09}
Dong~Hyun Jeong, Caroline Ziemkiewicz, Brian Fisher, William Ribarsky, and
  Remco Chang.
\newblock ipca: An interactive system for pca-based visual analytics.
\newblock {\em Computer Graphics Forum}, 28(3):767--774, 2009.

\bibitem{tyson2007science}
Will Tyson, Reginald Lee, Kathryn~M Borman, and Mary~Ann Hanson.
\newblock Science, technology, engineering, and mathematics (stem) pathways:
  High school science and math coursework and postsecondary degree attainment.
\newblock {\em Journal of Education for Students placed at risk},
  12(3):243--270, 2007.

\bibitem{XU2021100179}
Yongjun Xu, Xin Liu, Xin Cao, Changping Huang, Enke Liu, Sen Qian, Xingchen
  Liu, Yanjun Wu, Fengliang Dong, Cheng-Wei Qiu, Junjun Qiu, Keqin Hua, Wentao
  Su, Jian Wu, Huiyu Xu, Yong Han, Chenguang Fu, Zhigang Yin, Miao Liu, Ronald
  Roepman, Sabine Dietmann, Marko Virta, Fredrick Kengara, Ze~Zhang, Lifu
  Zhang, Taolan Zhao, Ji~Dai, Jialiang Yang, Liang Lan, Ming Luo, Zhaofeng Liu,
  Tao An, Bin Zhang, Xiao He, Shan Cong, Xiaohong Liu, Wei Zhang, James~P.
  Lewis, James~M. Tiedje, Qi~Wang, Zhulin An, Fei Wang, Libo Zhang, Tao Huang,
  Chuan Lu, Zhipeng Cai, Fang Wang, and Jiabao Zhang.
\newblock Artificial intelligence: A powerful paradigm for scientific research.
\newblock {\em The Innovation}, 2(4):100179, 2021.

\bibitem{you2019real}
Michael You and Jessica Yin.
\newblock Real-time visualization of neural network training to supplement
  machine learning education.
\newblock In {\em 2019 IEEE Integrated STEM Education Conference (ISEC)}, pages
  371--374. IEEE, 2019.

\bibitem{lamy2019visual}
Jean-Baptiste Lamy and Rosy Tsopra.
\newblock Visual explanation of simple neural networks using interactive
  rainbow boxes.
\newblock In {\em 2019 23rd International Conference Information Visualisation
  (IV)}, pages 50--55. IEEE, 2019.

\bibitem{Guidotti2018}
Riccardo Guidotti, Anna Monreale, Salvatore Ruggieri, Franco Turini, Fosca
  Giannotti, and Dino Pedreschi.
\newblock A survey of methods for explaining black box models.
\newblock {\em ACM Comput. Surv.}, 51(5), aug 2018.

\bibitem{Hao18}
Hao Li, Zheng Xu, Gavin Taylor, Christoph Studer, and Tom Goldstein.
\newblock Visualizing the loss landscape of neural nets.
\newblock In {\em Proceedings of the 32nd International Conference on Neural
  Information Processing Systems}, NIPS'18, page 6391–6401, Red Hook, NY,
  USA, 2018. Curran Associates Inc.

\bibitem{chatzimparmpas2020survey}
Angelos Chatzimparmpas, Rafael~M Martins, Ilir Jusufi, and Andreas Kerren.
\newblock A survey of surveys on the use of visualization for interpreting
  machine learning models.
\newblock {\em Information Visualization}, 19(3):207--233, 2020.

\bibitem{mariescu2019machine}
Radu Mariescu-Istodor and Ilkka Jormanainen.
\newblock Machine learning for high school students.
\newblock In {\em Proceedings of the 19th Koli calling international conference
  on computing education research}, pages 1--9, 2019.

\bibitem{kim2022development}
Jeongah Kim and Jaekwoun Shim.
\newblock Development of an ar-based ai education app for non-majors.
\newblock {\em IEEE Access}, 10:14149--14156, 2022.

\bibitem{rojas96neural}
Raul Rojas.
\newblock {\em Neural Networks - A Systematic Introduction}.
\newblock Springer-Verlag, Berlin, 1996.

\bibitem{Bengio2012}
Yoshua Bengio.
\newblock {\em Practical Recommendations for Gradient-Based Training of Deep
  Architectures}, pages 437--478.
\newblock Springer Berlin Heidelberg, Berlin, Heidelberg, 2012.

\bibitem{Chou19}
Chun-Nan Chou, Chuen-Kai Shie, Fu-Chieh Chang, Jocelyn Chang, and Edward~Y.
  Chang.
\newblock {\em Representation Learning on Large and Small Data}, chapter~1,
  pages 1--28.
\newblock John Wiley \& Sons, Ltd, 2019.

\bibitem{Lu20}
Lu~Lu, Yeonjong Shin, and George~Em Karniadakis.
\newblock Dying relu and initialization: Theory and numerical examples.
\newblock {\em Communications in Computational Physics}, 28(5):1671--1706,
  2020.

\bibitem{chandra2001parallel}
Rohit Chandra, Leo Dagum, David Kohr, Ramesh Menon, Dror Maydan, and Jeff
  McDonald.
\newblock {\em Parallel programming in OpenMP}.
\newblock Morgan kaufmann, 2001.

\bibitem{gonzalez2008digital}
Rafael~C. Gonzalez and Richard~E. Woods.
\newblock {\em Digital image processing}.
\newblock Prentice Hall, Upper Saddle River, N.J., 2008.

\bibitem{deng2012mnist}
Li~Deng.
\newblock The mnist database of handwritten digit images for machine learning
  research.
\newblock {\em IEEE Signal Processing Magazine}, 29(6):141--142, 2012.

\bibitem{DEPERLIOGLU2011392}
Omer Deperlioglu and Utku Kose.
\newblock An educational tool for artificial neural networks.
\newblock {\em Computers \& Electrical Engineering}, 37(3):392--402, 2011.

\bibitem{Maaten08}
Laurens van~der Maaten and Geoffrey Hinton.
\newblock Visualizing data using t-sne.
\newblock {\em Journal of Machine Learning Research}, 9(86):2579--2605, 2008.

\bibitem{mcinnes2018umap}
Leland McInnes, John Healy, and James Melville.
\newblock {UMAP}: Uniform manifold approximation and projection for dimension
  reduction.
\newblock {\em arXiv preprint arXiv:1802.03426}, 2018.

\end{thebibliography}

\end{document}